\documentclass[11pt, oneside]{article}   	
\usepackage{geometry}                		
\usepackage{graphicx}				
\usepackage{amssymb}
\usepackage{amsmath}
\usepackage{subfig}
\usepackage{color}
\usepackage{graphicx}

\usepackage{multirow}
\usepackage{adjustbox}
\usepackage{setspace}
\usepackage{natbib}
\setcitestyle{authoryear,open={(},close={)}}

\newtheorem{theorem}{Result}
\newtheorem{lemma}[theorem]{Lemma}

\newtheorem{assumption}[theorem]{Assumption}

\newcommand{\logit}{\operatorname{logit}}
\newcommand{\pr}{\operatorname{pr}}

\title{A coherent likelihood parametrization for doubly robust estimation of a causal effect with missing confounders}
\date{}							

\author{Katherine Evans, Isabel Fulcher, and Eric J. Tchetgen Tchetgen}

\begin{document}
	\maketitle
	\doublespacing

\begin{abstract}
Missing data and confounding are two problems researchers face in
observational studies for comparative effectiveness. \cite{Williamson_2012}
recently proposed a unified approach to handle both
issues concurrently using a multiply-robust (MR) methodology under
the assumption that confounders are missing at random. Their approach
considers a union of models in which any submodel has a parametric
component while the remaining models are unrestricted. We show that
while their estimating function is MR in theory, the possibility for
multiply robust inference is complicated by the fact that parametric
models for different components of the union model are not variation
independent and therefore the MR property is unlikely to hold in practice.
To address this, we propose an alternative transparent parametrization
of the likelihood function, which makes explicit the model dependencies
between various nuisance functions needed to evaluate the MR efficient
score. The proposed method is genuinely doubly-robust (DR) in that
it is consistent and asymptotic normal if one of two sets of modeling
assumptions holds. We evaluate the performance and doubly 
robust property of the DR method via a simulation study.
\end{abstract}

\section{Introduction}

Confounding bias and missing data are two major analytic challenges
in comparative effectiveness research using observational data such
as electronic medical records. While each problem has been thoroughly
studied separately, consolidated approaches for addressing both issues
are lacking. In the absence of missing data, confounding bias must
still be adjusted for in order to evaluate causal effects \citep{Hernan_et_al_2004}.
Researchers often use the g-formula for identifying the distribution
of counterfactuals from the observed data distribution \citep{Robins1986,Snowden2011}.
Inverse probability weighting estimators are commonly used and involve
modeling the propensity score \citep{RubinPS,Robins1986,hernan2000marginal}.
Doubly-robust estimators for causal effects have been well established
and widely studied \citep{Bang_and_Robins,LuncefordDavidian,robins2000,Vansteelandt2007}.
These estimators are doubly robust in the sense that they are consistent
and asymptotic normal if either the treatment mechanism (propensity
score) or outcome model is correctly specified, but not necessarily
both. These methods are also locally semiparametric efficient because
they achieve the semiparametric efficiency bound for the nonparametric
model when model misspecification is absent, that is, at the intersection
submodel of the union of specified models. 

Multiple imputation and inverse probability of censoring weighting
are increasingly popular methods for addressing missing data \citep{Rubin_MI,IPW1,IPW2}.
In the context of regression analysis, various weighting schemes to
account for missing covariates have been examined previously in the
literature \citep{Moore2009,RegrWeights,Parzen2002,TchetgenTchetgen2009}.
Semiparametric locally efficient methods are also available to address
data missing at random i.e., the probability of observing the full
data depends on the fully observed data only \citep{Kang2007,Bang_and_Robins}.
Robins, Rotnitzky and others examined improved augmented inverse weighted
estimators within the semiparametric framework \citep{RRZ,RR1,RR2,SRR}.
Additionally, \cite{Tsiatis2007} provides an extensive overview
of the state of the art for applying semiparametric theory to missing
data.

However, to date, few methods have considered joint inferences about
causal effects that are doubly or multiply robust in the
presence of missing data and confounding. This setting presents a
special challenge in that it involves the nesting of causal inference
in the missing data setting, each of which requires, to obtain the
parameter of interest, estimating a nuisance parameter while appropriately
accounting for the fact that nuisance parameters needed to adjust
for selection bias are entangled with nuisance parameters needed to
address confounding bias. Entangled in the sense that we now need to account 
for modeling both the confounder and the missingness of that confounder, which 
are both typically a nuisance in their own right. \cite{Davidian2005} 
presented a doubly-robust augmented inverse weighted estimator of
the causal effect of exposure when the outcome was missing. In their
2003 textbook, Robins and van der Laan give a unified theory for addressing
causal inference in the presence of missing data but do not address
specific challenges with identifying an appropriate parametrization
for the observed data when addressing both confounding adjustment
and incomplete confounder data \citep{Mark_Jamie_book}. This paper
addresses a special case of that general theory.

\cite{Williamson_2012} attempt to combine
existing methods in order to create a multiply-robust estimator. The
authors consider a union of four semiparametric models each of which
specifies parametric working models for either the missingness mechanism
or the missing covariates to account for missing data, and for either
the treatment mechanism or the outcome to account for confounding.
Multiply-robust estimation requires that each submodel of the union
model is a semiparametric model in that the correctly specified part
is parametric, while the remaining submodels are unrestricted. However,
we will show in this paper that the rest of the likelihood is, in
fact, restricted in at least one submodel of the union model and therefore
the multiply-robust property claimed by \cite{Williamson_2012} may
not be achievable in reality. An immediate implication of this phenomenon
is that in addition to possible lack of compatibility across submodels
of the union model, the intersection submodel of the union model may
in fact be empty. Therefore, unless one explicitly acknowledges the
overlap between components of the union model in the process of model
specification, one may in fact rule out the possibility of achieving
local efficiency.

In this paper we discuss the difficulty of achieving double robustness
in semiparametric missing data when full data nuisance parameters
are entangled with nuisance parameters needed to account for data
missing at random. We carefully examine the previously suggested multiply-robust
method and explain why it may fail to achieve the claimed multiply-robust
property. We then propose a solution that carefully identifies the
modeling assumptions through an alternative transparent parametrization
of the likelihood function, which makes explicit model dependencies
between various nuisance functions needed to evaluate the multiply-robust
estimating equation for the causal effect of interest. The proposed
method is genuinely doubly-robust in that it is consistent and asymptotically
normal if one of two sets of modeling assumptions holds. Further, due to the inherent
model dependencies, we establish that double-robustness to model misspecification is the 
best one hope to achieve in this setting. This paper suggests an approach that could easily be
adopted in other settings where one may wish to obtain a doubly-robust
estimator in the presence of entangled nuisance parameters. While
the paper focuses on the effect of treatment on the treated, the proposed
approach equally applies to the average causal effect.

\section{Preliminaries}

\subsection{Full Data Setting}

Let $A$ denote a binary treatment, $A\in\{0,1\}$, and let $Y$ be
the outcome in view with $Y_{1}$ and $Y_{0}$ denoting the potential
outcomes under treatment and control conditions respectively. Let
$W$ denote a set of pre-treatment covariates. The parameter of interest
is the effect of treatment on the treated on the additive scale, defined
as $E[Y_{1}-Y_{0}\mid A=1]=\theta-\Psi$ where $\theta=E[Y_{1}\mid A=1]$
and $\Psi=E[Y_{0}\mid A=1]$.

Throughout, we make the following standard causal assumptions in order
to identify the effect of treatment on the treated:
\begin{assumption}
\label{assumptionA}
Consistency: $Y=Y_{A}$ almost surely;
\end{assumption}
\begin{assumption}
\label{assumptionB}
No unmeasured confounding: $A\perp Y_{0}\mid W$;
\end{assumption}
\begin{assumption}
\label{assumptionC}
Positivity: $\frac{\pr(A=0\mid W)}{\pr(A=1\mid W)}>0$ almost surely.
\end{assumption}

Assumption 1 states that a person's observed outcome corresponds to
her potential outcome for the observed treatment. Assumption 2 states
that the treatment assignment is ignorable conditional on covariates
$W$, i.e. $W$ includes all common causes of $A$, $Y_{1}$ and $Y_{0}$. And assumption
3 states that there is no treated subject without an untreated counterpart.

Under assumption 1, $\theta=E[Y\mid A=1]$ \citep{Angrist,KennedyAndSmall}.
Under assumptions 1-3, $\Psi$ is well known to be non-parametrically
identified and is
\begin{eqnarray}
\Psi & = & E[Y_{0}\mid A=1]\nonumber \\
 & = & \frac{1}{\pr(A=1)}E\left[(1-A)\frac{\pr(A=1\mid W)}{\pr(A=0\mid W)}Y\right].\label{eq:Psi_full-1}
\end{eqnarray}

The following dual representation of (\ref{eq:Psi_full-1}) is also
of interest
\begin{eqnarray*}
\Psi & = & E\left[E[Y\mid A=0,W]\mid A=1\right]\\
 & = & \int f(w\mid A=1)\int yf(y\mid A=0,w)d\mu\left(w,y\right)
\end{eqnarray*}

where $\mu$ is a dominating measure of the distribution of $\left(W,Y\right)$.

Any regular and asymptotically linear estimator $\hat{\Psi}$ of $\Psi$
satisfies
\begin{eqnarray*}
n^{\frac{1}{2}}\left(\hat{\Psi}-\Psi\right) & = & n^{-\frac{1}{2}}\sum_{i=1}^{n}\iota\left(A_{i},W_{i},Y_{i};\Psi\right)+o_{p}\left(1\right)
\end{eqnarray*}
where $\iota\left(A_{i},W_{i},Y_{i};\Psi\right)$ is a zero-mean function,
called the $i^{th}$ influence function for $\Psi$. The influence
function characterizes the behavior of the estimator (such as the
asymptotic distribution) and under certain condition, may also be
used to define an estimating equation to obtain an estimator with
the corresponding influence function. For functionals defined on nonparametric
models, as will be considered in this paper, there exists a unique
influence function that is semiparametric efficient under the nonparametric
model. Influence functions were first introduced by Huber \citep{Huber1972}
in the context of robust statistics and later developed in semiparametric
theory, in the sense of  \cite{Bickel1993}. In
either context, influence functions represent the influence of a single
observation on the estimator.

The efficient influence function of $\Psi$ in the nonparametric model
in which assumptions 1-3 hold, but the form of the observed data likelihood is
unrestricted is \citep{Hahn1998}{\small{},
\begin{eqnarray}
\iota_{Full}(\Psi) & = & \frac{I(A=0)}{\pr(A=1)}\frac{f(A=1\mid W)}{f(A=0\mid W)}(Y-E[Y\mid A=0,W])+\frac{I(A=1)}{\pr(A=1)}(E[Y\mid A=0,W]-\Psi).\label{eq:IF_small}
\end{eqnarray}
}{\small \par}

In order to use the efficient influence function as an estimating
function for $\Psi$ when, as is typically the case in observational
studies, $W$ is high dimensional, one must estimate the nuisance
functions $f(A\vert W)$ and $E[Y\mid A=0,W]$ using low dimensional
parametric working models. The solution to the resulting estimating equation
is doubly robust for $\Psi$ in that it is consistent provided we
consistently estimate the propensity score, $f(A\mid W)$, or the
outcome model, $f\left(Y\mid A=0,W\right)$, but not necessarily both.
Additionally, the estimator achieves the nonparametric efficiency
bound in the absence of model misspecification. If interest instead lies in the average causal effect, the assumptions can be slightly modified to obtain identification results from the existing literature. The remaining results will equally apply.

\subsection{Missing Data Setting}

Next, suppose that only a subset of covariates, $C$, of $W$ is fully
observed while $L$ is missing for a subset of participants where
$W=\{L,C\}$. Therefore the observed data can be written as $(Y,A,C,RL,R)$,
where $R$ is an indicator function which is equal to 1 when $L$
is observed and is otherwise equal to 0. Define $O=(Y,A,C)$, the
fully observed data. Furthermore, suppose that $L$ is missing at
random. The data now presents a non-monotone missingness pattern with
respect to missing confounders and missing counterfactual outcomes.

In order to address missing data, we make the following additional
assumptions:
\begin{assumption}
\label{assumptionD}
$\pi=\pr(R=1\mid A,L,C,Y)>0$ almost surely;
\end{assumption}
\begin{assumption}
\label{assumptionE}
Conditional exchangeability: $L\perp R\mid A,C,Y$.
\end{assumption}

Assumption 4 is a positivity assumption and states that there is a
positive probability of observing any possible value of $(A,C,L,Y)$
in the complete cases. Assumption 5 is a missing at random assumption
and states that the conditional distribution of $L$ given $A,C,Y$
is the same in incomplete and complete cases.

Under assumptions 1-5, the efficient influence function of $\Psi$
in the nonparametric model in which the observed data distribution
is unrestricted is
\begin{eqnarray}
\iota_{Miss}(\Psi) & = & \frac{R}{\pi}\iota_{Full}(\Psi)-(\frac{R}{\pi}-1)E[\iota_{Full}(\Psi)\mid O].\label{eq:if_miss}
\end{eqnarray}

The efficient influence function in equation (\ref{eq:if_miss}) depends
on the following functions: the propensity score, $p=\pr(A=1\mid L,C)$,
the outcome model, $m=m\left(Y\mid A,L,C\right)$, the missing data
mechanism, $\pi=\pr(R=1\mid A,Y,C)$, and the density of $L$ given
$A$, $C$ and $Y$, $t=t(L\mid A,C,Y)$.

The efficient influence function is appealing as a basis for obtaining
inferences about $\Psi$, mainly because of the following multiple
robust property:
\begin{eqnarray}
E\left[\iota_{Miss}(p,m,\pi,t;\Psi)\right] & = & 0\label{eq:E(IF_miss)=00003D0}
\end{eqnarray}
if $\Psi$ is evaluated at the truth, and one of the following statements
hold:

(i) $p$ and $\pi$ are evaluated at the truth;

(ii) $m$ and $\pi$ are evaluated at the truth;

(iii) $p$ and $t$ are evaluated at the truth;

(iv) $m$ and $t$ are evaluated at the truth.

Additionally, at the intersection submodel where all of the models
are evaluated at the truth, the variance of $\iota_{Miss}(\Psi)$
achieves the semiparametric efficiency bound for the union of models
(i)-(iv) at the intersection submodel.

A closely related multiply-robust property of the efficient influence
function to account for missing confounders was established for the
average causal effect by \cite{Williamson_2012}.

However, because in practice one must estimate $p$, $m$, $\pi$,
and $t$ under corresponding low dimensional working models, model
incompatibility may render the multiply-robust property given above
infeasible, as we show next. The approach considers four submodels,
$p$, $m$, $\pi$, and $t$ and four unions of those submodels. These
submodels are semiparametric in the sense that in practice, within
each submodel of the union model, two models are parametrically specified,
but the remaining two are not modeled and left unrestricted. However,
this cannot hold as the specified models in at least one of the submodels
of the union model places restrictions on components of the likelihood
not explicitly modeled in the submodel. When, as is typically the
case in practice, each component of the likelihood is eventually modeled
separately, conflict may arise in two separate models for the same
component of the likelihood, therefore ruling out the possibility
for multiple robustness and local efficiency. To explain how this
potential conflict in model specification may arise, consider that
the joint likelihood of all four models is $\left[f\left(Y,A,L,C\right)\right]^{R}\left\{ \int f\left(Y,A,l,C\right)dl\right\} ^{1-R}f\left(R\mid Y,A,L,C\right)$.
The submodels $t(L\mid A,C,Y)$ and $m(Y\mid A,C,L)$ are not variation
independent, as they both encode an association between $Y$ and $L$
given $A$ and $C$. Similarly, the models for $t(L\mid A,C,Y)$ and
$p=\pr(A=1\mid L,C)$ are not variation independent because both densities
encode the association between $A$ and $L$ given $C$. Because these
various functions are not variation independent, a choice of model
for one may restrict modeling options for the other.  As a result of the lack of variation independence, multiply-robust cannot be achieved under a coherent parametrization of the likelihood which acknowledges
the model dependence revealed above. Furthermore, unless one such
parametrization can be established, local efficiency may also not
be attainable because the intersection submodel may be empty in presence
of conflicting models. We have provided a straightforward illustration of this phenomenon in the
supplementary materials. The following section provides a coherent parametrization 
of the observed data likelihood under
which a certain degree of double robustness can be achieved and local
efficiency remains a genuine possibility.

\section{Reparametrization}

We propose one possible parametrization of the conditional likelihood function, $f(L,Y\mid A,C)$, which makes explicit the model dependencies between
nuisance functions that are needed to evaluate the efficient score
given by (\ref{eq:E(IF_miss)=00003D0}) in order to obtain an estimator,
as described later in section 4. The proposed approach is based on
a conditional odds ratio symmetric parametrization of
a joint conditional distribution.

Following \cite{Chen_2007} and \cite{DR_OddsRatio} we define the
conditional odds ratio function of $A$ and $Y$ given $L$ as
\begin{eqnarray*}
\chi\left(A,Y\vert L\right) & = & \frac{f\left(A\mid Y,L\right)f\left(a_{0}\mid y_{0},L\right)}{f\left(a_{0}\mid Y,L\right)f\left(A\mid y_{0},L\right)}
\end{eqnarray*}
where $\left(a_{0},y_{0}\right)$ is a reference value.

\cite{Chen_2007} established that the joint distribution
of $A$ and $Y$ given $L$ can be written as
\begin{eqnarray*}
g\left(A,Y\mid L\right) & = & \frac{\chi\left(A,Y\mid L\right)f\left(A\mid y_{0},L\right)f\left(Y\mid a_{0},L\right)}{\int\int\chi\left(a,y\mid L\right)f\left(a\mid y_{0},L\right)f\left(y\mid a_{0},L\right)d\mu\left(a,y\right)},
\end{eqnarray*}
where $\int\int\chi\left(a,y\mid L\right)f\left(a\mid y_{0},L\right)f\left(y\mid a_{0},L\right)d\mu\left(a,y\right)<\infty$.
This parametrization is attractive because $\chi\left(A,Y\mid L\right)$,
$f\left(A\mid y_{0},L\right)$, and $f\left(Y\mid a_{0},L\right)$
are variation independent in that the choice of a parametric model
for one component does not restrict available model choices for another
and their joint parameter space is the product space of their respective
parameter spaces. We repeatedly make use of the variation independent parameterization result from \cite{Chen_2007} in the supplementary materials to prove the following result.

\setcounter{theorem}{0}

\begin{theorem}
\label{theorem1}

Let $f=f(L,Y\mid A,C)$ be the distribution of $L$ and $Y$ given
$A$ and $C$ where $f(L\mid a_0,C)$, $f(Y\mid l_0,A,C)$, $\chi(L,Y\mid A,C)$,
and $\chi\left(A,L\mid C\right)$ are variation independent parameters. Then, $f$ can be written as\emph{\small{}
\begin{eqnarray*}
f & = & \frac{\chi(L,Y\mid A,C)f(Y\mid l_0,A,C)}{K\left(A,C\right)}\frac{f(L\mid a_0,C)\chi\left(A,L\mid C\right)}{f(l_0\mid a_0,C)\int\chi(L,y\mid A,C)f(y\mid A,C,l_0)d\mu(y)},
\end{eqnarray*}
}
\noindent where $K\left(A,C\right)=\frac{\int\int\chi\left(l,y\mid A,C\right)f\left(l\mid y_{0},A,C\right)f\left(y\mid l_{0},A,C\right)d\mu\left(l,y\right)}{f(l_0\mid y_0,A,C)}$ and $(a_0,l_0,y_0)$ are reference values.\\
\end{theorem}
This theorem gives a variation independent parameterization of the likelihood and makes explicit that to model $f$ we must model $\chi(L,Y\mid A,C)$, $\chi\left(A,L\mid C\right)$, $f(L\mid a_0,C)$,
and $f(Y\mid l_0,A,C)$. Similarly we can parameterize the propensity score, $p$ as
\begin{eqnarray*}
f(A\mid L,C) & = & \frac{\chi(A,L\mid C)f(A\mid l_0,C)}{\tilde{K}(C)},
\end{eqnarray*}
where $\tilde{K}(C)=\sum_{a}P(A=a\mid l_0,C)\chi(a,L\mid C)$. This
reparametrization makes explicit the fact that the propensity score
can be expressed in terms of $\chi(A,L\mid C)$ and $f(A\mid l_0,C)$,
which are variation independent. 

Therefore, both $f$ and $p$ require the same specification of $\chi(A,L\mid C)$, so that they are not variation independent.
Furthermore both $m$ and $t$ share the odds ratio of $L$ and $Y$
given $C$ and $A$. This implies that assuming a submodel for $m$
also places a restriction on the submodel for $t$ which cannot remain
unrestricted as assumed by \cite{Williamson_2012} in their claim
to achieve multiple robustness.

\section{Doubly Robust Inference}

Let $j=j\left(L\mid A=0,C\right)$ denote the distribution of $L$
given $A=0$ and $C$ and let $j\left(\alpha\right)=j\left(L\mid A=0,C;\alpha\right)$
be a parametric model for $j$. Define $W=W\left(Y,L\mid A,C\right)$
such that $W\left(0,L\mid A,C\right)=W\left(Y,0\mid A,C\right)=1$
and $W\ge0$ so that $W$ is the true conditional odds
ratio function for $Y$ and $L$ given $A$ and $C$ \citep{Chen_2007}
and let $w\left(\omega\right)=w\left(Y,L\mid A,C;\omega\right)$ be
a parametric model for $w$. Let $\chi\left(\beta\right)=\chi\left(A,L\mid C,\beta\right)$
be a parametric model for $\chi(A,L\mid C)$ and let $\pi\left(\eta\right)=\pr(R=1\mid A,C,Y;\eta)$
be a parametric model for $\pi$. Let $r=r\left(Y\mid A,L=0,C\right)$
be a model for the distribution of $Y$ given $A$, $L=0$, and $C$
and let $r\left(\theta\right)=r\left(Y\mid A,L=0,C;\theta\right)$
be a parametric model for $r$. Additionally, let $h=\pr(A=1\mid L=0,C)$
and let $h\left(\kappa\right)=\pr(A=1\mid L=0,C;\kappa)$ be a parametric
model for $h$. 

An estimator $\left(\hat{\alpha},\hat{\omega},\hat{\beta},\hat{\theta}\right)$
of the parameters $\left(\alpha,\omega,\beta,\theta\right)$, can
be found by using direct likelihood maximization of the observed data.
This entails maximizing the observed data likelihood, 
$$\prod\left[f\left(Y,L,\mid A,C;\alpha,\omega,\beta,\theta\right)\right]^{R}\left[\int f\left(Y,l,\mid A,C;\alpha,\omega,\beta,\theta\right)d\mu(l)\right]^{1-R}$$
An estimator of $\eta$, $\hat{\eta}$, can be found by fitting $\pi\left(\eta\right)$
to the observed data 
$$\hat{\eta}=\arg\max_{\eta}\left[\sum R_{i}\log\pi(\eta)+\left(n-\sum R_{i}\right)\log\left(1-\pi(\eta)\right)\right]$$
where $n$ is the total number of subjects. Finally, $\kappa$ can
be estimated using inverse probability weighting using $1/\pi\left(\hat{\eta}\right)$
as weights in the complete cases.

\begin{theorem}
\label{theorem2}

Define $\hat{\Psi}$ as the solution to
\begin{eqnarray*}
\mathbb{P}_{n}\left(\iota_{Miss}\left(\hat{\Psi};\hat{\alpha},\hat{\omega},\hat{\beta},\hat{\theta},\hat{\eta},\hat{\kappa}\right)\right) & = & 0,
\end{eqnarray*}
where $\mathbb{P}_{n}(.)=\frac{1}{n}\sum_{i}(.)_{i}$ and where $\iota_{Miss}\left(\hat{\Psi};\hat{\alpha},\hat{\omega},\hat{\beta},\hat{\theta},\hat{\eta},\hat{\kappa}\right)$
is equal to $\iota_{Miss}\left(\Psi;\alpha,\omega,\beta,\theta,\eta,\kappa\right)$
evaluated at $\left(\hat{\alpha},\hat{\omega},\hat{\beta},\hat{\theta},\hat{\eta},\hat{\kappa}\right)$.
Then under standard regularity conditions, $\hat{\Psi}$ is consistent
and asymptotically normal if $\chi\left(\beta\right)$ is correctly
specified and in addition either (i) $\pi\left(\hat{\eta}\right)$
and $h\left(\hat{\kappa}\right)$ are consistent for $\pi$ and $h$
or (ii) $j\left(\hat{\alpha}\right)$, $w\left(\hat{\omega}\right)$,
and $r\left(\hat{\theta}\right)$ are consistent for $j$, $w$, and
$r$. Additionally, at the intersection submodel where all of the
models are evaluated at the truth, the variance of $\hat{\Psi}$ achieves
the semiparametric efficiency bound for the union of models (i) and
(ii).
\end{theorem}

An alternative approach using a more standard parametrization can
sometimes be used, provided that the parametrization can be shown
to satisfy the variation dependence described in Theorem 1, which ensures
the existence of a joint distribution for $\left(L,A,Y\mid C\right)$.
A standard parametrization in this case implies specifying parametric
models for $p$, $t$, $\pi$, and $m$ needed to evaluate the efficient
influence function (\ref{eq:if_miss}). 

Let $p\left(\lambda\right)$ be a parametric model for $p$, $t\left(\phi\right)$
be a parametric model for $t$, $\pi\left(\eta\right)$ be a parametric
model for $\pi$, and $m\left(\nu\right)$ be a parametric model for
$m$. An estimator of $\eta$,$\hat{\eta}$ , can be found by fitting
$\pi\left(\eta\right)$ on the observed data by using, using, for
example, a logistic regression of $R$ on $A$, $C$, and $Y$. We
can estimate $\lambda$ by using inverse probability of censoring
weighting using $\frac{1}{\pi\left(\hat{\eta}\right)}$ as weights
in the complete cases. For example we might fit a weighted logistic
regression of $A$ on $C$, and $L$. By specifying $m$ and $t$
as normal with constant variance, one may ensure the existence of
a corresponding joint distribution of $\left(Y,L\right)$ given $A$
and $C$, provided the mean of $Y$ given $L$, $A$, and $C$ is
linear in $L$ and likewise the mean model for $L$ given $Y$, $A$,
and $C$ is linear in $Y$. Assumption 5, missing at random, allows
us to estimate $t\left(\phi\right)$ using the complete cases by using,
for example, a standard linear regression of $L$ on $A$, $C$, and
$Y$. Finally, we describe a simple Monte Carlo algorithm to estimate
$m\left(\hat{\nu}\right)$:
\begin{enumerate}
\item Create $M$ duplicates of the data.
\item Where $L$ is missing, ``fill in'' the missing variable with a random
draw from $t\left(\hat{\phi}\right)$ .
\item Stack all $M$ datasets in long format and estimate $m\left(\hat{\nu}\right)$.
In practice this will involve fitting a standard model for $Y$on
$A$, $L$, and $C$. For example, a standard main effects linear
model.
\end{enumerate}
We then have the following result.

\begin{theorem}
\label{theorem3}

Define $\hat{\Psi}$ as the solution to
\begin{eqnarray*}
\mathbb{P}_{n}\left(\iota_{Miss}\left(\hat{\Psi};\hat{\lambda},\hat{\nu},\hat{\eta},\hat{\phi}\right)\right) & = & 0.
\end{eqnarray*}
 Then under standard regularity conditions, $\hat{\Psi}$ is consistent
and asymptotically normal if the implied form of $\chi\left(\beta\right)$
is correctly specified and in addition either (i) $\pi\left(\hat{\eta}\right)$
and $p\left(\hat{\lambda}\right)$ are consistent for $\pi$ and $p$
or (ii) $m\left(\hat{\nu}\right)$ and $t\left(\hat{\phi}\right)$
are consistent for $m$ and $t$, but not necessarily both.\textbf{
}As before, at the intersection submodel where all of the models are
evaluated at the truth, the variance of $\hat{\Psi}$ achieves the
semiparametric efficiency bound for the union of models (i) and (ii).
\end{theorem}

Then, it is straightforward to show that the solution to equation
(\ref{eq:if_miss}) is
\begin{eqnarray}
\hat{\Psi} & = & \mathbb{P}_{n}\Bigg\{\frac{R}{\hat{\pi}}\left\{ \frac{I(A=0)}{\hat{\pr}(A=1)}\frac{\hat{p}}{1-\hat{p}}(Y-\hat{\mu}_{Y}^{0})+\frac{I(A=1)}{\hat{\pr}(A=1)}\hat{\mu}_{Y}^{0}\right\} \label{eq: psi hat}\\
 &  & -\mbox{\,}(\frac{R}{\hat{\pi}}-1)\left\{ \frac{I(A=0)}{\hat{\pr}(A=1)}YE\left[\frac{\hat{p}}{1-\hat{p}}\mid Y,A=0,C\right]\right\} \nonumber \\
 &  & +\mbox{\,}(\frac{R}{\hat{\pi}}-1)\left\{ \frac{I(A=0)}{\hat{\pr}(A=1)}E\left[\frac{\hat{p}}{1-\hat{p}}\hat{\mu}_{Y}^{0}\mid Y,A=0,C\right]\right\} \nonumber \\
 &  & -\mbox{\,}(\frac{R}{\hat{\pi}}-1)\left\{ \frac{I(A=1)}{\hat{\pr}(A=1)}E\left[\hat{\mu}_{Y}^{0}\mid Y,A=1,C\right]\right\} \Bigg\},\nonumber 
\end{eqnarray}
where $\hat{\mu}_{Y}^{0}\left(\nu\right)=E[Y\mid A=0,L,C;\hat{\nu}]$,
$\hat{\pi}=\pi\left(\hat{\eta}\right)$, and $\hat{p}=p\left(\hat{\lambda}\right)$.

The asymptotic distribution of the estimator can be found as follows.
Let $Q_{R}\left(\hat{\eta}\right)$ be an individual contribution
to the score for $\eta$, $Q_{A}\left(\hat{\lambda}\right)$ be an
individual contribution to the score for $\lambda$, $Q_{L}\left(\hat{\phi}\right)$
be an individual contribution to the score for $\phi$, and $Q_{Y}\left(\hat{\nu}\right)$
be an individual contribution to the score for $\nu$. For example,
\begin{eqnarray*}
Q_{R}(\eta) & = & \frac{d}{d\eta}\log\left[\pi(\eta)^{R}\left(1-\pi(\eta)\right)^{1-R}\right].
\end{eqnarray*}

Also let $Z\left(\hat{\Psi},\hat{\lambda},\hat{\nu},\hat{\eta},\hat{\phi}\right)$
be an individual contribution to the estimating equation for $\Psi$.
Let $\Xi=\left(\eta,\lambda,\phi,\nu\right)$ and defin{\footnotesize{}e
\begin{eqnarray*}
Q(\hat{\Xi}) & = & \left(\begin{array}{c}
Q_{R}\left(\hat{\eta}\right)\\
Q_{A}\left(\hat{\lambda}\right)\\
Q_{L}\left(\hat{\phi}\right)\\
Q_{Y}\left(\hat{\nu}\right)
\end{array}\right).
\end{eqnarray*}
}{\footnotesize \par}

Then, under standard regularity conditions,
\begin{eqnarray*}
n^{\frac{1}{2}}(\hat{\Psi}-\Psi) & = & n^{-\frac{1}{2}}E\left[\frac{dZ}{d\Psi}\right]^{-1}\sum_{i=1}^{n}\left\{ Z(\Psi,\Xi)-\frac{d}{d\Xi}E\left[Z\left(\Psi,\Xi\right)\right]E\left[\frac{dQ}{d\Xi}\right]^{-1}Q\left(\Xi\right)\right\} +op\left(1\right).
\end{eqnarray*}

Therefore, a consistent estimator of the asymptotic variance of $\sqrt{n}(\hat{\Psi}-\Psi)$
is
\[
\left[\mathbb{P}_{n}\frac{dZ}{d\hat{\Psi}}\right]^{-1}\mathbb{P}_{n}\left[V\left(\hat{\Psi},\hat{\Xi}\right)V^{T}\left(\hat{\Psi},\hat{\Xi}\right)\right]\left[\mathbb{P}_{n}\left(\frac{dZ}{d\hat{\Psi}}\right)^{T}\right]^{-1}
\]
where
\begin{eqnarray*}
V\left(\hat{\Psi},\hat{\Xi}\right) & = & Z(\Psi,\Xi)-\frac{d}{d\Xi}E\left[Z\left(\Psi,\Xi\right)\right]E\left[\frac{dQ}{d\Xi}\right]^{-1}Q\left(\Xi\right)
\end{eqnarray*}
with all marginal expectations replaced by their empirical counterparts.

Alternatively, we recommend using the nonparametric bootstrap to obtain
estimates of the variance. 

In their application to the B-Aware trial, 
\cite{Williamson_2012} use models that are compatible with this new
parametrization. As a result, the estimating equation under those
choices of models is doubly-robust, though not multiply-robust as
they claim. However, their simulation models do not satisfy our proposed
parametrization and therefore fail to be compatible, thus there is
no chance of it being even doubly-robust. Moreover, the favorable
simulation results obtained by the authors can be explained by two
reasons. The first reason is that the effect of the missing confounder on the 
 exposure was small compared to the fully observed confounders in the model. 
The second reason for their favorable simulation results is that the
model for the missing confounder is only mildly misspecified. In order
to misspecify the model for the missing confounder, the authors omit
variables with small regression coefficients and therefore little
influence in the model while retaining variables with larger coefficients
\citep{Williamson_2012}. Further, their complete case estimator performs 
well which is indicative of a favorable data setting.

\section{Simulation Study}

We report a simulation study comparing finite sample performance of
our doubly-robust estimator to a number of existing methods. We compared
our doubly-robust estimator to an estimator that used Monte Carlo
direct likelihood maximization, one using inverse probability of censoring
weights, as well as a complete case estimator, a naive estimator that
drops the missing confounder, and an estimator calculated from the
complete dataset where $L$ was observed for all subjects. The last
estimator is obviously not feasible in the presence of missing data
however provides a benchmark to assess efficiency loss due to missing
data. 

In the first set of simulations, we simulated $C$ by summing draws
from a standard normal distribution and a uniformly distributed variable on the interval $\left(-1,1\right)$. The treatment, $A$, was Bernoulli with $\pr(A=1\mid C;\zeta)=p_{A}=\zeta_{0}+\zeta_{1}C$.
For this simulation we chose $(\zeta_{0},\zeta_{1})=(-0.44,0.40)$.
The outcome, $Y$, was chosen to be normal conditional on $A$ and
$C$, with $Y=\upsilon_{0}+\upsilon_{1}A+\upsilon_{2}C+\epsilon_{y}$
where $\epsilon_{Y}\sim N(0,\sigma_{Y}^{2})$, $(\upsilon_{0},\upsilon_{1},\upsilon_{2},\sigma_{Y}^{2})=(0.2,0.38,0.3,0.51)$.
Similarly, $L$ was chosen to be normal conditional on $A$ and $C$,
with $L=\alpha_{0}+\alpha_{1}A+\alpha_{2}C+\epsilon_{L}$ where $\epsilon_{L}\sim N(0,\sigma_{L}^{2})$
and such that $Cov(\epsilon_{Y},\epsilon_{L})=\sigma_{YL}$ and $(\alpha_{0},\alpha_{1},\alpha_{2},\sigma_{L}^{2},\sigma_{YL})=(-0.15,0.215,0.14,0.43,0.21)$.
As a consequence, the distribution of $L$ given $A$, $Y$ and $C$,
$t\left(\phi\right)$, was normal such that $E[L\mid A,Y,C]=\phi_{0}+\phi_{1}A+\phi_{2}Y+\phi_{3}C=\mu_{L}$
where $(\phi_{0},\phi_{1},\phi_{2},\phi_{3})=(-0.23,0.058,0.41,0.016)$,
the distribution of $Y$ given $C$, $L$ and $A$, $m\left(\nu\right)$,
was normal such that $E[Y\mid A,L,C]=\nu_{0}+\nu_{1}A+\nu_{2}L+\nu_{3}C=\mu_{Y}$
where $(\nu_{0},\nu_{1},\nu_{2},\nu_{3})=(0.27,0.275,0.49,0.23)$,
and $p\left(\lambda\right)=\logit[\pr(A=1\mid L,C;\lambda)]=\lambda_{0}+\lambda_{1}L+\lambda_{2}C$
was the propensity score with $(\lambda_{0},\lambda_{1},\lambda_{2})=(-0.42,0.5,0.36)$.
These models appropriately encode the variation dependence described
in Result 1 and ensure the existence of a joint distribution of $\left(L,A,Y\mid C\right)$.
These simulations were used for the first 6 figures below (\emph{$a$-$f$})
and have only a moderate relationship between $L$ and $C$. This
setting is especially useful in order to explore the potential impact
of model misspecification of the propensity score which will be explained
further below.

In the second set of simulations, $C$ was generated as in the previous
simulation. The treatment $A$ was Bernoulli with $\pr(A=1\mid C)=p_{A}=\zeta_{0}+\zeta_{1}C$
as above. However, for this simulation we chose $(\zeta_{0},\zeta_{1})=(-0.44,0.38)$.
The outcome, $Y$, was chosen to be Normal conditional on $A$ and
$C$ as above. $L$ was chosen to be Normal conditional on $A$ and
$C$ similarly to the previous simulation but instead with $\alpha_{2}=0.914$
in order to have a strong relationship between $L$ and $C$. As a
consequence, $t\left(\phi\right)$ was Normal such that $E[L\mid A,Y,C]=\phi_{0}+\phi_{1}A+\phi_{2}Y+\phi_{3}C=\mu_{L}$
where $(\phi_{0},\phi_{1},\phi_{2},\phi_{3})=(-0.23,0.058,0.41,0.79)$,
$m\left(\nu\right)$ was Normal such that $E[Y\mid A,L,C]=\nu_{0}+\nu_{1}A+\nu_{2}L+\nu_{3}C=\mu_{Y}$
where $(\nu_{0},\nu_{1},\nu_{2},\nu_{3})=(0.27,0.275,0.49,-0.146)$
and $p\left(\lambda\right)=\logit[\pr(A=1\mid L,C;\lambda)]=\lambda_{0}+\lambda_{1}L+\lambda_{2}C$
was the propensity score with $(\lambda_{0},\lambda_{1},\lambda_{2})=(-0.42,0.5,0.10)$.
These simulations were used for the final 2 figures below ($g$ and
$h$) and have a strong relationship between $L$ and $C$. This setting
is useful in order to explore the impact of model misspecification
of the joint distribution of $Y$ and $L$, which will be explained
further below.

In both simulations, $R$ was $Bernoulli\left(\pi\right)$ with $\pi\left(\eta\right)=\logit\left[\pr(R=1\mid A,C,Y)\right]=\eta_{0}+\eta_{1}A+\eta_{2}C+\eta_{3}Y$
where $\left(\eta_{0},\eta_{1},\eta_{2},\eta_{3}\right)=\left(1,-1.75,-1.75,1.25\right)$.
In both simulations, on average, $\pr(R=1)\approx0.61$. The observed
data were therefore $n=2,500$ realizations of $\left(R,RL,Y,A,C\right)$.
Many more details concerning the simulation can be found in the supplementary materials. 

In simulations, we implemented the following estimators for comparison: standard inverse probability of censoring weights (IPCW) estimation, Monte Carlo direct likelihood maximization (MCDLM) using 100 imputed datasets, complete-case analysis (CC), and a naive estimator (Naive) that drops the missing confounder $L$ completely and evaluates (\ref{eq:Psi_full-1}) upon substituting an estimate of $\pr(A=1\mid C)$ for $p$.

For the various methods we fitted the following models. For the missingness
mechanism $\pi\left(\eta\right)$, we fitted a logistic regression.
Similarly, we fitted a logistic regression for the propensity score,
$p\left(\lambda\right)$, using inverse probability weighting with
$\frac{1}{\pi(\hat{\eta})}$ as weights in the complete cases. For
the distribution of the missing variable, $t\left(\phi\right)$, we
fitted a main effects linear model. Finally, for the outcome model,
$m\left(\nu\right)$, we fitted a main effects linear model.

The inverse probability of censoring weights estimator required $\pi\left(\eta\right)$
as well as $p\left(\lambda\right)$. The Monte Carlo direct likelihood
maximization estimator used $t\left(\phi\right)$ as well as $p\left(\lambda\right)$.
The complete case estimator only required $p\left(\lambda\right)$.
The Naive estimator required a logistic regression for $A$ with main
effects for $C$ alone, $\tilde{p}\left(\tilde{\lambda}\right)=\pr\left(A=1\mid C;\tilde{\lambda}\right)$.
All these methods were compared to the proposed doubly-robust estimator
which required $p\left(\lambda\right)$, $\pi\left(\eta\right)$,
$t\left(\phi\right)$, and $m\left(\nu\right)$.

For the complete-case, naive, Monte Carlo direct likelihood maximization, and 
inverse probability of censoring weights,
we calculated the effect of treatment on the treated for each method
using equation (\ref{eq:Psi_full-1}) for $\Psi$. For the naive estimator
the odds, $p\left(\lambda\right)/\left[1-p\left(\lambda\right)\right]$,
were replaced with $\tilde{p}\left(\tilde{\lambda}\right)/\left[1-\tilde{p}\left(\tilde{\lambda}\right)\right]$
and for the inverse probability of censoring weights estimator the
odds were estimated with inverse probability weighting. Our proposed
estimator was calculated using equation (\ref{eq: psi hat}).

The misspecified versions of each model were as follows. The missingness
mechanism, $\pi$, was misspecified by only using $C$ in the regression,
$\pi^{*}=\pr(R=1\mid C;\lambda^{*})$. In order to misspecify a model
for $p$ or $f$ we simply (incorrectly) set the coefficient on $C$
to $0$ in the working model. This form of misspecification was chosen
in order to preserve the structure of the odds ratio between $A$
and $L$ given $C$, $\chi(A,L\mid C)$, wherever it is required as
seen in Section 3. 

If $L$ and $C$ are strongly correlated, particularly when the coefficient
on $C$ in the propensity score model, $\lambda_{2}$, is small, then
not including $C$ in the propensity score will not be far off from
the truth as $L$ will likely suffice to account for confounding.
However if $L$ and $C$ are weakly correlated, then any imputation
of $L$ that sets the coefficient on $C$ to zero will not be far
off from the true model that includes $C$. Therefore we impose a
weak correlation for the simulations exploring misspecification of
$p$ and a strong correlation for those misspecifying $f$ as described
above. For settings where both are misspecified, we impose a weak
correlation $L$ and $C$. We denote the incorrect propensity score
as $p^{*}$ and the incorrect joint distribution of $Y$ and $L$
given $A$ and $C$ as $f^{*}$.

\begin{figure}
	\centering
	\subfloat[All models correct]{ 
		\includegraphics[scale=.12]{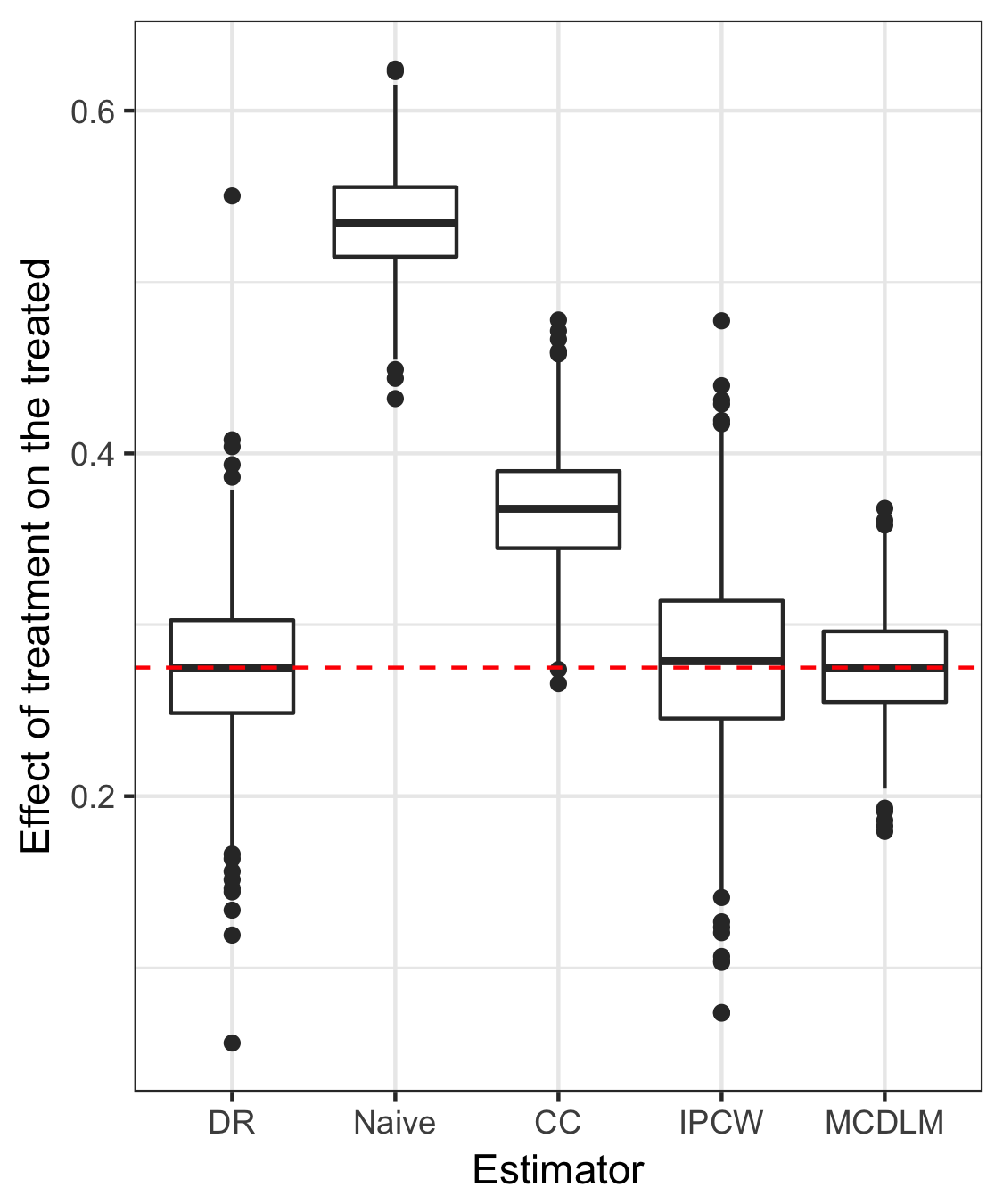} }
	\subfloat[Model for $f$ incorrect]{
	\includegraphics[scale=.12]{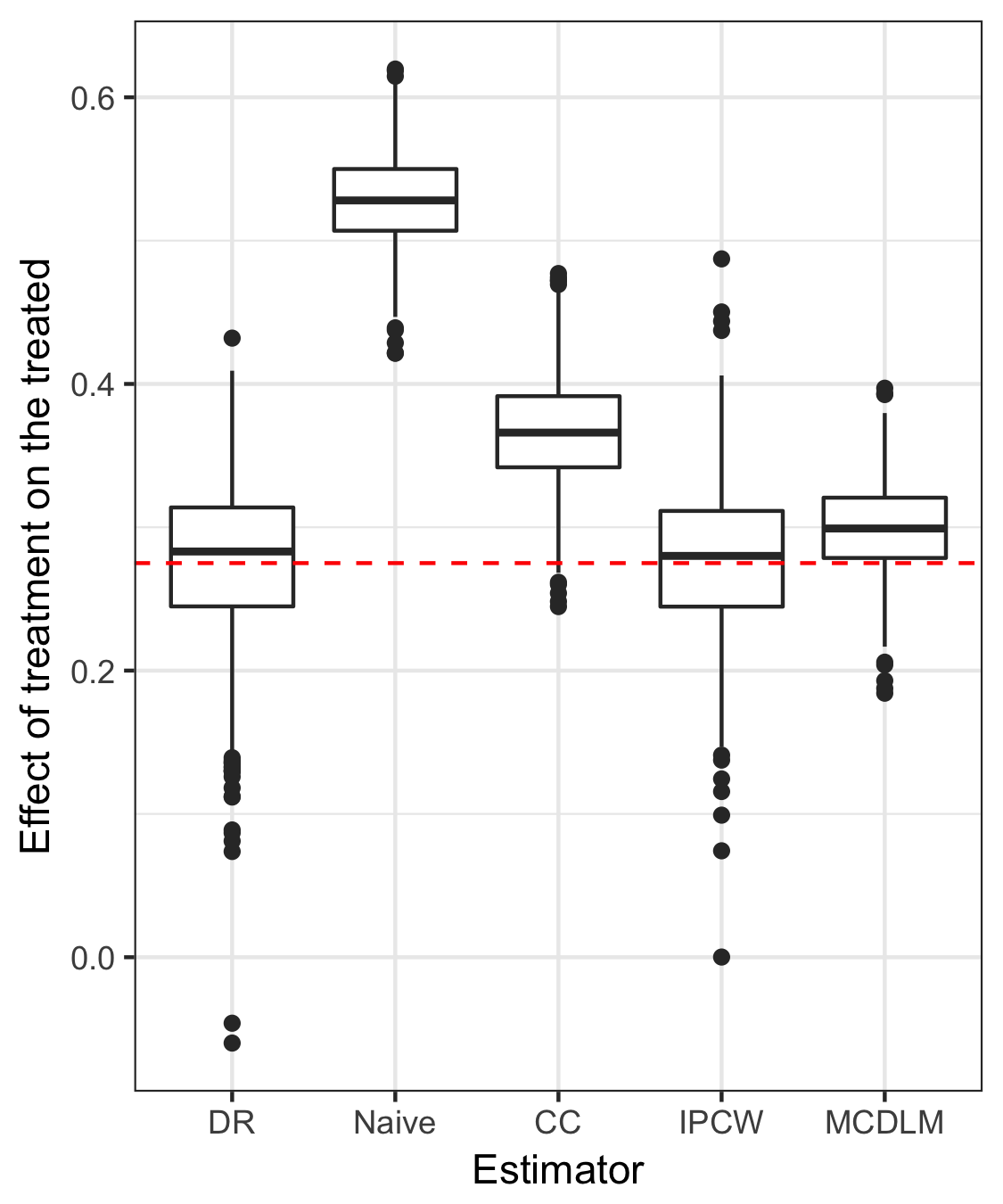}}
	\subfloat[Model for $p$ incorrect]{
	\includegraphics[scale=.12]{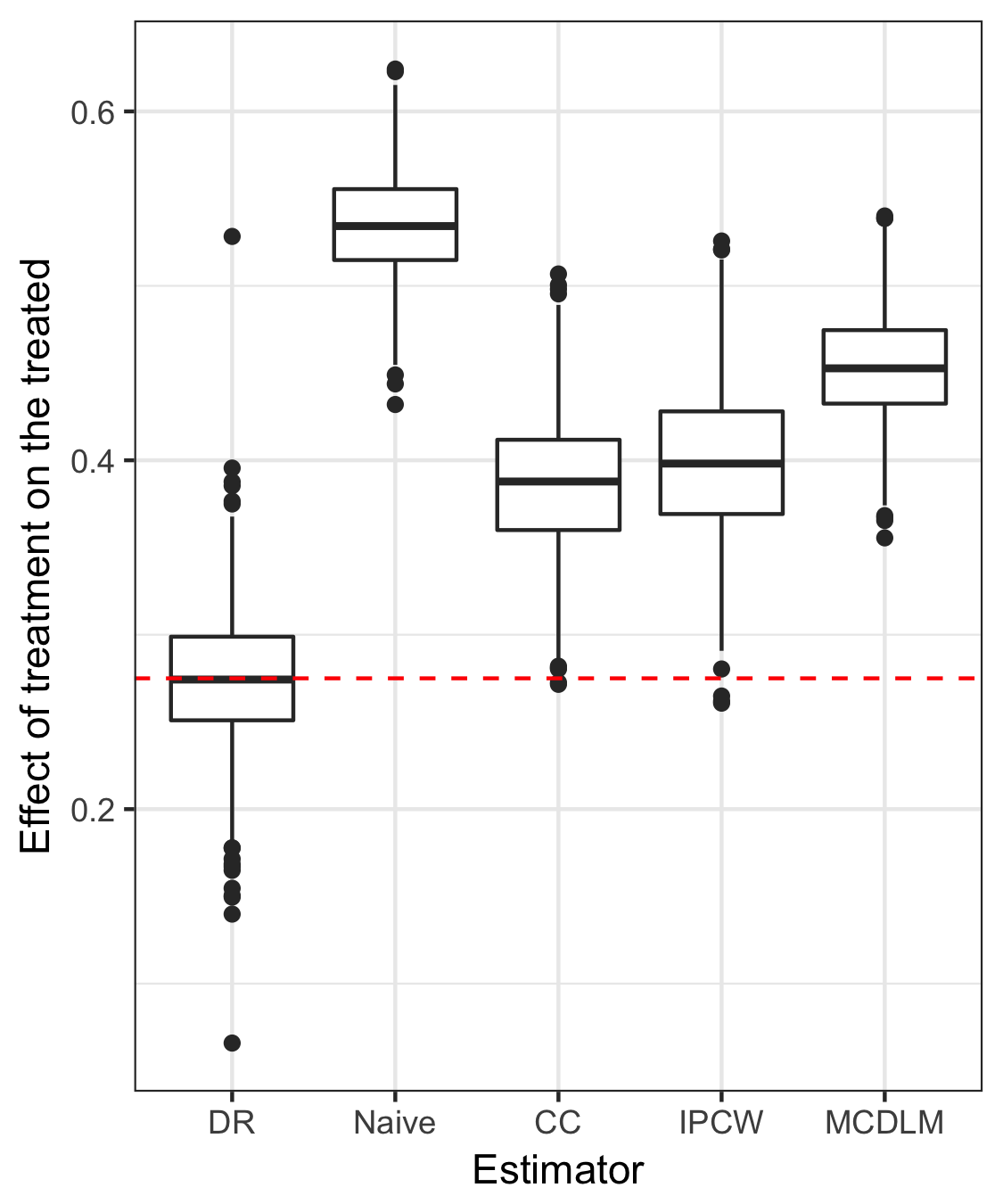}} \\
\subfloat[Models for $\pi$ incorrect]{ 
	\includegraphics[scale=.12]{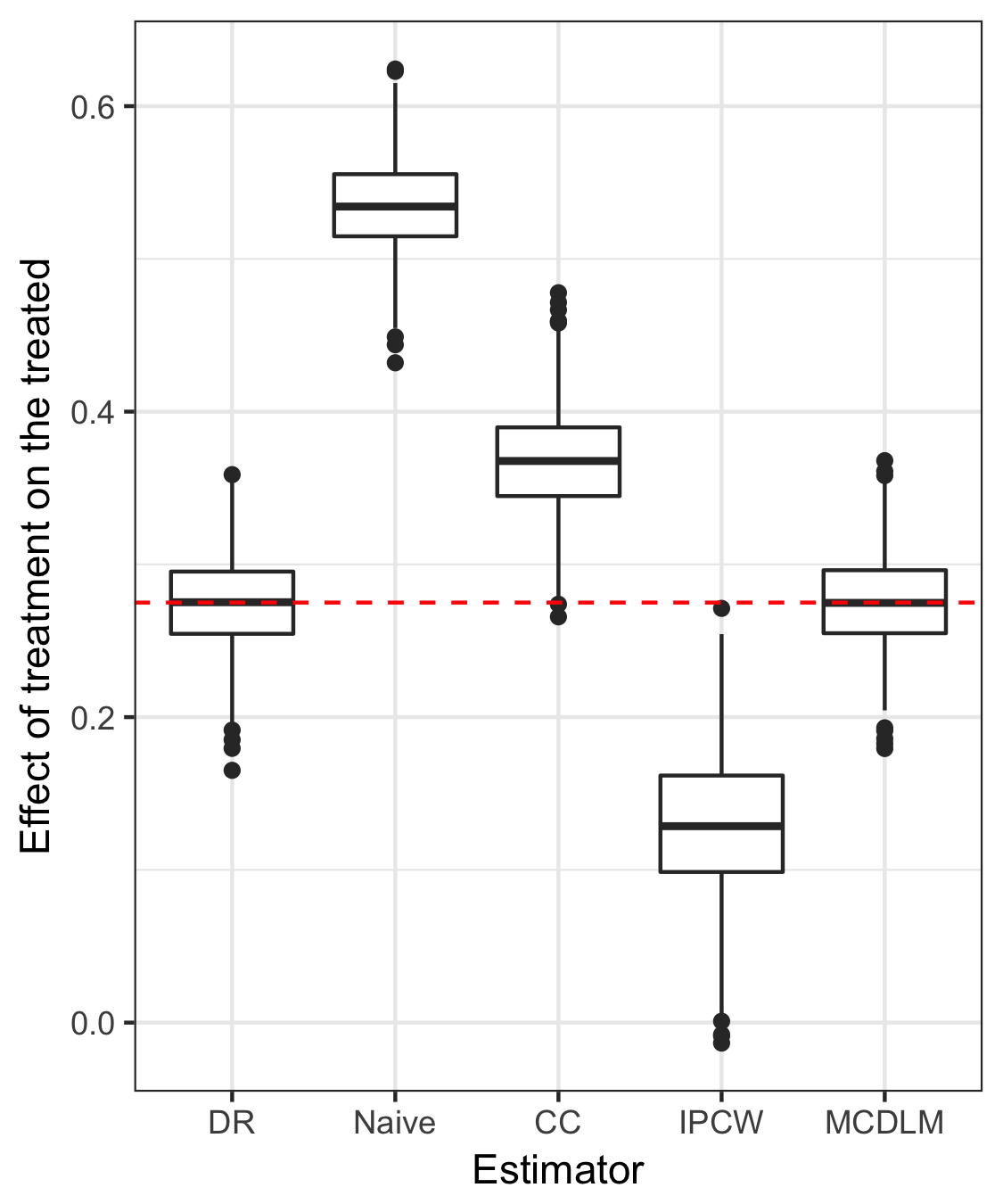} }
\subfloat[Models for $\pi$ and $p$ incorrect]{
	\includegraphics[scale=.12]{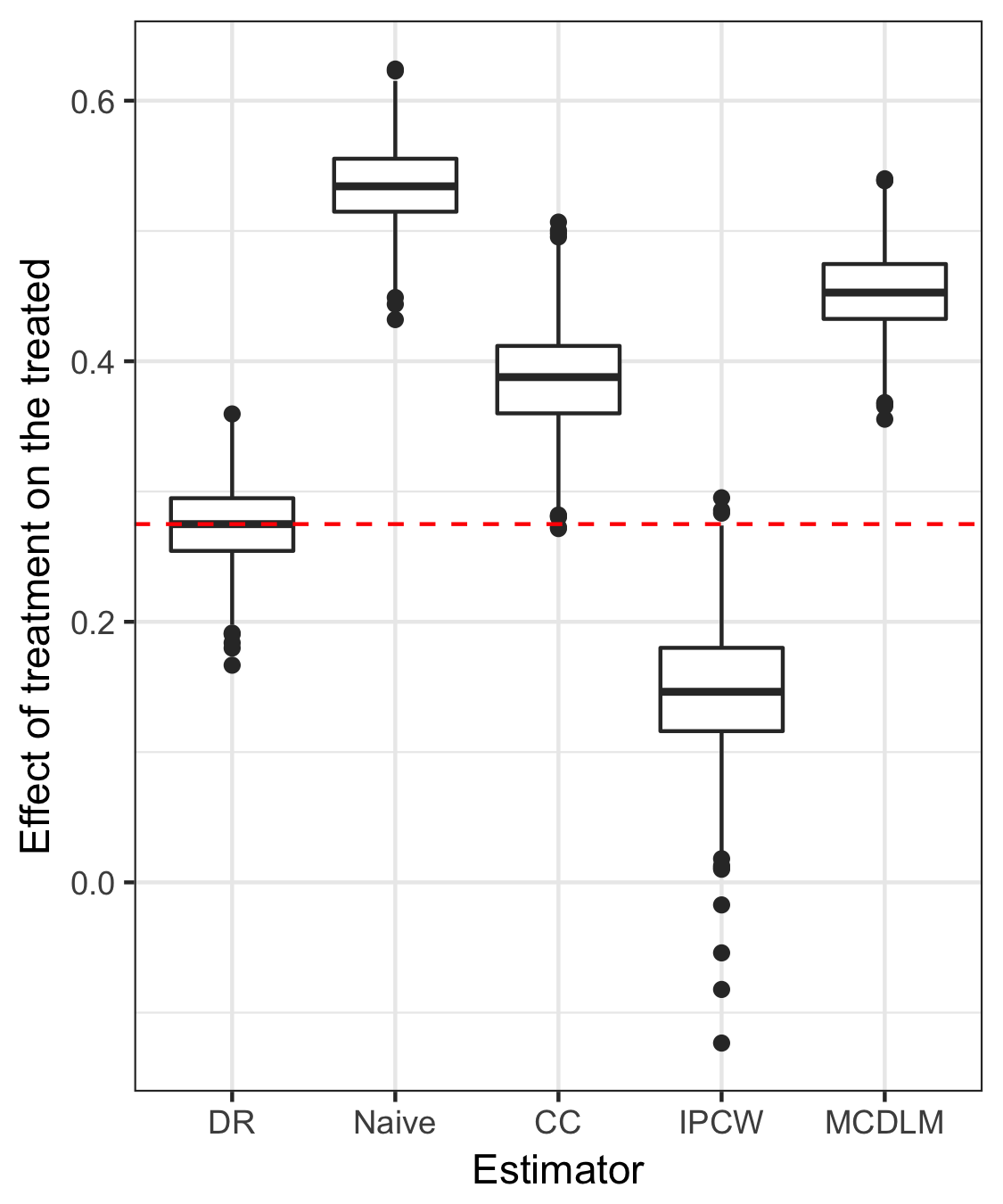}}
\subfloat[Models for $f$ and $p$ incorrect]{
	\includegraphics[scale=.12]{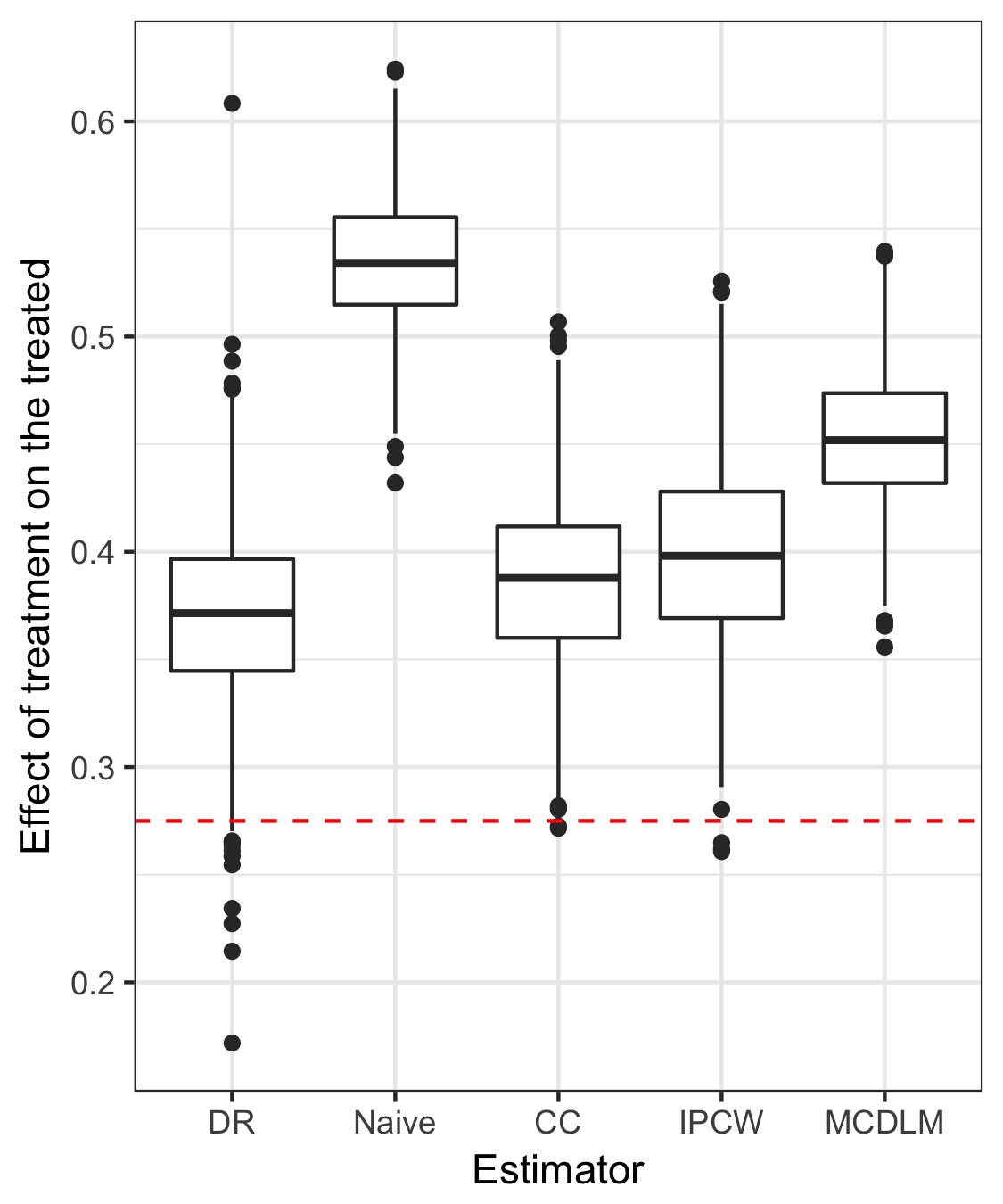}} \\
\subfloat[Models for $f$ and $\pi$ incorrect]{
	\includegraphics[scale=.12]{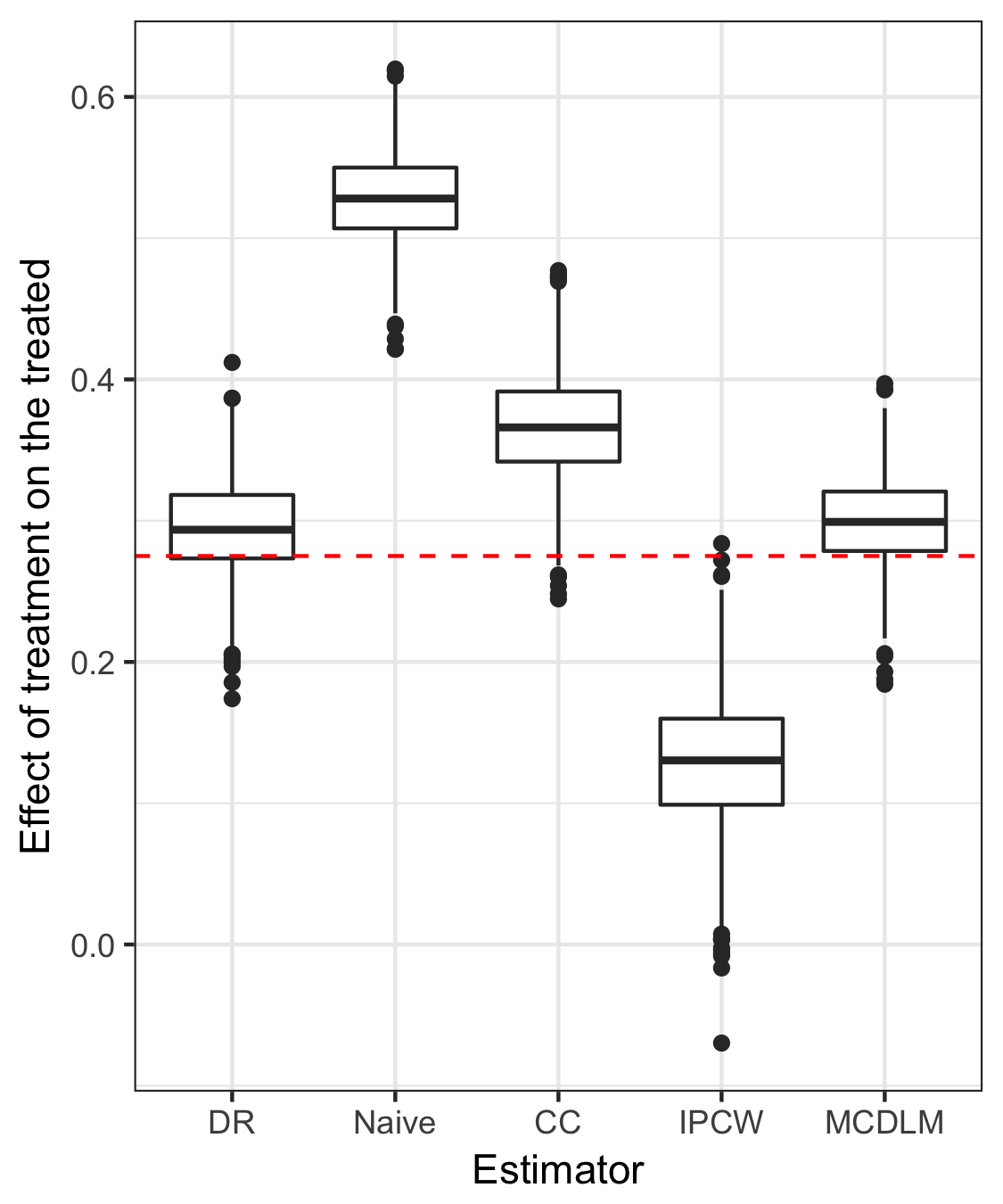}}
\subfloat[Models for $f$, $p$, and $\pi$ incorrect]{
	\includegraphics[scale=.12]{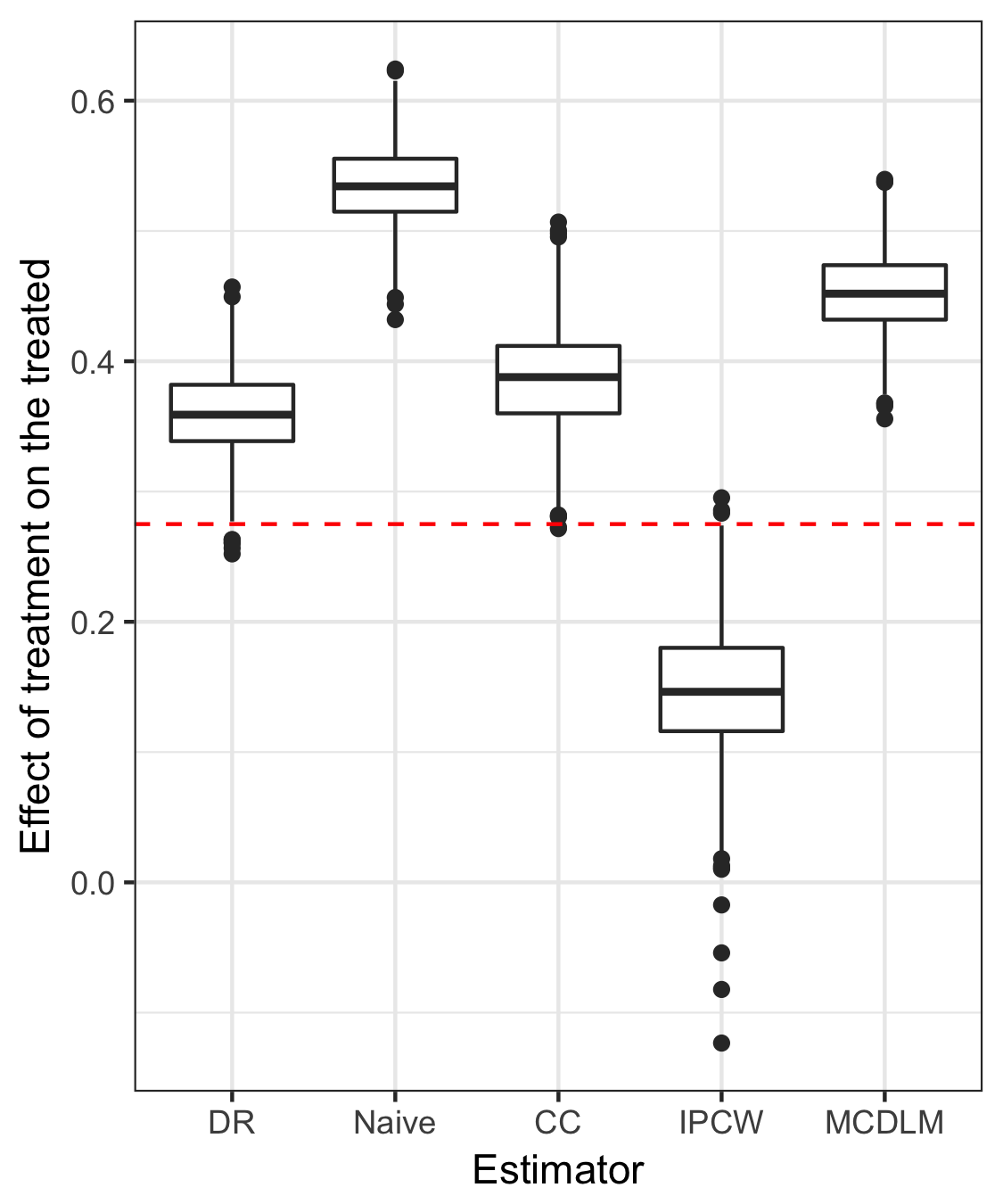}} \\
		\caption{Simulation results for various model misspecifications in $f$, $p$, and $\pi$ across various estimators where the red line indicates the truth. DR is our proposed doubly-robust estimator, Naive is the naive estimator that drops a missing confounder, CC is the complete-case estimator, IPCW is the inverse probability of censoring weights estimator, and MCDLM is the Monte Carlo direct likelihood maximazation estimator.}
		\label{fig:P2Simulation-Results}
	\end{figure}

Figure \ref{fig:P2Simulation-Results} summarizes the results in the
form of Monte Carlo boxplots for the estimated population effect of
treatment on the treated for $1,000$ Monte Carlo samples of $2,500$
subjects for each of the following scenarios: (a) all models were
correctly specified, (b) $f^{*}$ used in place of $f$, (c) $p^{*}$
used in place of $p$, (d) $\pi^{*}$ used in place of $\pi$, 
(e) $\pi^{*}$ and $p^{*}$ used in place of $\pi$ and
$p$  (f) $f^{*}$ and $p^{*}$ used in place of $f$ and
$p$, (g) $f^{*}$ and $\pi^{*}$ used in place of $f$ and $\pi$, and (h) $f^{*}$, $\pi^{*}$, and $p^{*}$ used in place of $f$,$\pi$,
and $p$. 

Regardless of model misspecification, the naive and complete-case
estimators are biased ($a$-$g$) as expected. Similarly, the inverse probability of censoring weights estimator is biased when $p^{*}$ or $\pi^{*}$ are used in place of $p$ or
$\pi$ (\textbf{$c$}-$h$) as it requires both and not a model
for $f$. Additionally, the inverse probability of censoring weights
estimator tended to have large variance compared to the other estimators,
even under correct specification for $p$ and $\pi$. The Monte Carlo
direct likelihood maximization estimator is biased when $p^{*}$ is
used in place of $p$($c$,$e$, $f$, $h$). When $f^{*}$ is used in place of
$f$, but $p$ is correctly used ($b$, $g$) the Monte Carlo direct
likelihood maximization estimator is biased, but not overly so. This
is likely an artifact of the simulation design regarding the correlation
between $L$ and $C$ as explained above. Finally, our doubly-robust
estimator is only biased under the settings we expected, namely when
$f$ is misspecified along with $p$ or $\pi$ or both misspecified
($f$-$h$). Even in settings where the doubly-robust estimator
is biased, the bias is less than that of the other biased estimators.
In the setting where $\pi^{*}$ and $f^{*}$ are used in place of
$\pi$ and $f$, the bias is comparable to the bias in the
Monte Carlo direct likelihood maximization estimator. However, this
may be an artifact of the simulation design. Overall,
despite a few anomalies, the simulations are in line with expectations.
The simulation used in \cite{Williamson_2012}
did not allow for the range of settings we have explored. Furthermore, their model for
$R$ did not include $Y$. As a result, their missing data mechanism
assumption was stronger than missing at random and fairly mild such
that their complete case estimates had little or no bias.

\section{Discussion}

Analysts are commonly faced with missing data when using observational
data to estimate a causal effect. This is particularly true in the
setting of two stage non-monotone missingness, such as when potential
confounding information is missing along with counterfactual outcomes.
The difficulty arises when full data nuisance parameters are entangled
with nuisance parameters which are needed to account for data missing
at random. In such a setting it is unlikely that researchers will
know the underlying mechanisms for the missingness and confounding.
Therefore, model misspecification is a likely source of bias when
using standard statistical analysis methods. In this paper we have
explained why the proposed method of \cite{Williamson_2012}
fails to achieve the claimed multiply-robust property by carefully
examining model dependencies. We identified the modeling assumptions
through an alternative parametrization of the joint distribution of
the outcome and missing confounder in order to understand the nuisance
parameter entanglements. In this paper we propose a coherent likelihood
parametrization and an estimator of the effect of treatment on the
treated that accounts for both missingness and potential confounding
and that is robust to partial model misspecification.

The simulation study supported the conclusion that
 our proposed estimator is doubly robust and outperformed existing methods but still failed to be multiply
robust as we argued on theoretical basis. Moreover, we only considered
a setting in which a single confounder had missing data. It is more
common that several variables may be missing possibly in arbitrary
patterns across individuals \citep{MissingData,RRZ,BaoLuo_and_Eric}.
Therefore it would be important to extend our approach to allow for
arbitrary missing data patterns.

\bibliography{mybibarxiv}

\newpage

\begin{center}
\Huge{\textit{Supplementary Materials}}
\end{center}

\setcounter{theorem}{0}

\section*{Influence Function Derivation in Full Data Setting} 

\begin{assumption}
	Consistency: $Y=Y_{A}$ almost surely;
\end{assumption}
\begin{assumption}
	No unmeasured confounding: $A\perp Y_{0}\mid W$, where $W=(C,L)$;
\end{assumption}

Then, $\Psi=E[Y_{0}\mid A=1]=E\left[E[Y\mid A=0,W]\mid A=1\right]=\int f(w\mid A=1)\int yf(y\mid A=0,w)d\mu\left(w,y\right)$.

Consider a function of the observed data, $O$ , $F_{t}(O)$ such
that $F_{0}(O)=F(O)$. Then $\Psi_{t}=\Psi(F_{t})=\int f_{t}(w\mid A=1)\int yf_{t}(y\mid A=0,w)d\mu\left(w,y\right)$.

Our goal is to write $\frac{d\Psi_{t}}{dt}$ as $E[\iota_{Full}\times S(O)]$
where $\iota_{Full}$ is the influence function and $S(O)$ is the
score function where
\begin{eqnarray*}
	S(O) & = & \frac{d}{dt}\log f_{t}(O)\\
	& = & \frac{d}{dt}\left[\log f_{t}(Y\mid A,W)+\log f_{t}(A\mid W)+\log f_{t}(W)\right]\\
	& = & \frac{d}{dt}\left[\log f_{t}(Y\mid A,W)+\log f_{t}(W\mid A)+\log f_{t}(A)\right].
\end{eqnarray*}

Then,
\begin{eqnarray*}
	\frac{d\Psi_{t}}{dt} & = & T_{1}+T_{2}
\end{eqnarray*}
where $T_{1}=\int\int\frac{d}{dt}yf_{t}(y\mid A=0,w)f(w\mid A=1)d\mu\left(w,y\right)$
and $T_{2}=\int\int\frac{d}{dt}f_{t}(w\mid A=1)yf(y\mid A=0,w)d\mu\left(w,y\right)$.
We will drop the subscript for ease of notation. 

Then,
\begin{eqnarray*}
	T_{1} & = & \int\int\frac{d}{dt}yf_{t}(y\mid A=0,w)(w\mid A=1)yd\mu\left(w,y\right)\\
	& = & \int\int\int y\frac{I(A=0)}{f(a\mid w)}\frac{d}{dt}\frac{f_{t}(y\mid a,w)}{f(y\mid a,w)}\frac{f(w\vert A=1)}{f(w)}\\
	& \times & f(y\mid a,w)f(a\mid w)f(w)d\mu\left(w,a,y\right)\\
	& = & \int\int\int y\frac{I(A=0)}{f(A=0\mid w)}S(y\mid a,w)\frac{f(w\mid A=1)}{f(w)}\frac{\pr(A=1)}{\pr(A=1)}f(O)d\mu\left(w,a,y\right)\\
	& = & \int\int\int y\frac{I(A=0)}{f(A=0\mid w)}\frac{f(w,A=1)}{f(w)\pr(A=1)}S(y\mid a,w)f(O)d\mu\left(w,a,y\right)\\
	& = & \int\int\int y\frac{I(A=0)}{f(A=0\mid w)}\frac{f(A=1\mid w)f(w)}{f(w)\pr(A=1)}S(y\mid a,w)f(O)d\mu\left(w,a,y\right)\\
	& = & \int\int\int y\frac{I(A=0)}{f(A=0\mid w)}\frac{f(A=1\mid w)}{\pr(A=1)}S(y\mid a,w)f(O)d\mu\left(w,a,y\right)\\
	& = & \int\int\int(y-E[Y\mid A=0,w])\frac{I(A=0)}{f(A=0\mid w)}\frac{f(A=1\mid w)}{\pr(A=1)}\\
	& \times & f(O)[S(y\mid a,w)+S(a\mid w)+S(w)]d\mu\left(w,a,y\right).
\end{eqnarray*}

Therefore, the influence function for $T_{1}$ is
\begin{eqnarray*}
	\iota_{1} & = & (Y-E[Y\mid A=0,W])\frac{I(A=0)}{f(A=0\mid W)}\frac{f(A=1\mid W)}{\pr(A=1)}.
\end{eqnarray*}

Next
\begin{eqnarray*}
	T_{2} & = & \int\int\frac{d}{dt}f_{t}(w\mid A=1)yf(y\mid A=0,w)d\mu\left(w,y\right)\\
	& = & \int\int\frac{d}{dt}\frac{f_{t}(w\mid A)}{f(w\mid A)}E[Y\mid A=0,w]\frac{I(A=1)}{f(A)}f(A)f(W\mid A)d\mu\left(w,y\right)\\
	& = & \int\int S(w\mid A)f(w\mid A)f(A)E[y\mid A=0,w]\frac{I(A=1)}{f(A)}d\mu\left(w,y\right)\\
	& = & \int\int\int\left[S(y\mid w,a)+S(w\mid a)\right]f(y\mid w,a)f(w\mid a)f(a)E[Y\mid A=0,w]\frac{I(A=1)}{f(a)}d\mu\left(w,a,y\right)\\
	& = & \int\int\int\left[S(y\mid w,a)+S(w\mid a)+S(a)\right]f(O)\frac{I(A=1)}{f(a)}\left\{ E[Y\mid A=0,w]-\Psi\right\} d\mu\left(w,a,y\right).
\end{eqnarray*}

Therefore the influence function for $T_{2}$ is
\begin{eqnarray*}
	\iota_{2} & = & (E[Y\mid A=0,W]-\Psi)\frac{I(A=1)}{f(A)}.
\end{eqnarray*}

Now we can see that
\begin{eqnarray*}
	\frac{d\Psi_{t}}{dt} & = & E[\iota_{Full}\times S(O)]
\end{eqnarray*}

where
\begin{eqnarray*}
	\iota_{Full}(O) & = & \iota_{1}+\iota_{2}\\
	& = & \frac{I(A=0)}{\pr(A=1)}\frac{\pr(A=1\mid W)}{\pr(A=0\mid W)}(Y-E[Y\mid A=0,W])+\frac{I(A=1)}{\pr(A=1)}(E[Y\mid A=0,W]-\Psi).
\end{eqnarray*}
If one were to use the efficient influence function as an estimating
equation for $\Psi$, then
\begin{eqnarray*}
	\Psi & = & \frac{1}{\pr(A=1)}E\left[(1-A)\frac{\pr(A=1\mid W)}{\pr(A=0\mid W)}Y\right].
\end{eqnarray*}

\section*{Proof of Double Robustness for Full Data Setting}

If one were to use the efficient influence function,
as an estimating equation for $\Psi$, one would need to estimate
the nuisance functions $p(A\mid W)$ and $m\left(Y\mid A=0,L,C\right)$.
The resulting estimator, $\hat{\Psi}$, is double robust for $\Psi$
in that it will be consistent provided we correctly specify a model
for the propensity score, $p(A\mid W)$, or the outcome model, $m\left(Y\mid A=0,L,C\right)$,
but not necessarily both. To show this property, consider that $\hat{\Psi}$
will be consistent if $E[\iota]=0$ with the expectation taken at
the true value of $\Psi$. We demonstrate this property is true is
either $p(A\mid W)$ or $m\left(Y\mid A=0,L,C\right)$ is correct. 

If $m\left(Y\mid A=0,L,C\right)$ is correct and letting $p^{*}(A=1\mid W)$
denote the incorrect propensity score, then 
\begin{eqnarray*}
	E[\iota_{Full}] & = & E\left[\frac{I(A=0)}{\pr(A=1)}\frac{p^{*}(A=1\mid W)}{p*(A=0\mid W)}(Y-E[Y\mid A=0,W])+\frac{I(A=1)}{\pr(A=1)}(E[Y\mid A=0,W]-\Psi)\mid A,W\right]\\
	& = & E\left[\frac{I(A=0)}{\pr(A=1)}\frac{p^{*}(A=1\mid W)}{p^{*}(A=0\mid W)}\left(E[Y\mid A=0,W]-E[Y\mid A=0,W]\right)\right]\\
	&  & +\mbox{ }E\left[\frac{I(A=1)}{\pr(A=1)}(E[Y\mid A=0,W]-\Psi)\mid A,W\right]\\
	& = & E\left[\frac{I(A=1)}{\pr(A=1)}(E[Y\mid A=0,W]-\Psi)\right]\\
	& = & E\left[\frac{I(A=1)}{\pr(A=1)}(E\left[E[Y\mid A=0,W]\mid A=1\right]-\Psi)\right]\\
	& = & E\left[\frac{I(A=1)}{\pr(A=1)}(\Psi-\Psi)\right]\\
	& = & 0.
\end{eqnarray*}

Now let $p(A=1\mid W)=p\left(W\right)$ for ease of notation and suppose
$p\left(W\right)$ is correct while letting $E^{*}[Y\mid A=0,W]=b^{*}\left(W\right)$
denote the incorrect outcome expectation, then
\begin{eqnarray*}
	E[\iota_{Full}\mid W] & = & E\left[\frac{I(A=0)}{\pr(A=1)}\frac{p(W)}{1-p(W)}\left(Y-b^{*}\left(W\right)\right)+\frac{I(A=1)}{f(A=1)}\left(b^{*}\left(W\right)-\Psi\right)\mid Y,A,W\right]\\
	& = & \frac{1}{\pr(A=1)}E\left[(1-A)\frac{p(W)}{1-p(W)}\left(E[Y\mid A=0]-b^{*}\left(W\right)\right)+A\left(b^{*}\left(W\right)-\Psi\right)\mid Y,A,W\right]\\
	& = & \frac{1}{\pr(A=1)}E\left[p(W)\left(E[Y\mid A=0]-b^{*}\left(W\right)\right)+p(W)\left(b^{*}\left(W\right)-\Psi\right)\mid Y,A,W\right]\\
	& = & \frac{1}{\pr(A=1)}E\left[p(W)(E[Y\mid A=0]-\Psi)\mid Y,A,W\right]\\
	& = & \frac{1}{\pr(A=1)}E\left[p(W)(E[Y\mid A=0,W]-\Psi)\right]\\
	& = & \frac{1}{\pr(A=1)}E\left[p(W)(E\left[E[Y\mid A=0,W]\vert A=1\right]-\Psi)\right]\\
	& = & \frac{1}{\pr(A=1)}E\left[\Psi-\Psi\right]\\
	& = & 0.
\end{eqnarray*}

\section*{Influence Function Derivation for Missing Data Setting}

Recall that
\begin{eqnarray*}
	\iota_{Full}(O) & = & \frac{I(A=0)}{\pr(A=1)}\frac{f(A=1\mid W)}{f(A=0\mid W)}(Y-E[Y\mid A=0,W])+\frac{I(A=1)}{f(A=1)}(E[Y\mid A=0,W]-\Psi).
\end{eqnarray*}

Then the influence function for the missing data problem is
\begin{eqnarray*}
	\iota_{Miss} & = & \frac{R}{\pi}\iota_{Full}(O)-(\frac{R}{\pi}-1)E[\iota_{Full}(O)\mid O]\\
	& = & \frac{R}{\pi}\left(\iota_{Full}(O)-E[\iota_{Full}(O)\mid O]\right)+E[\iota_{Full}(O)\mid O],
\end{eqnarray*}
where $\pi=\pr(R=1\mid A,Y,C)$.

We, in theory, need to correctly specify:

$p=\pr(A=1\mid W)$ or $m=m(Y\mid A,C,L)$

and

$\pi=\pr(R=1\mid A,Y,C)$ or $t=t(L\mid A,C,Y)$.

If $t$ is correctly specified $E[\iota_{Full}(O)]=E\left[E[\iota_{Full}(O)\mid O]\right]$
and that if $p$ or $m$ are correct then $E[\iota_{Full}(O)]=0$
and in turn, $E[\iota_{Full}(O)]=E\left[E[\iota_{Full}(O)\mid O]\right]$.

From the above expressions for $\iota_{Miss}$, let:
\begin{eqnarray*}
	U_{1} & = & \frac{R}{\pi}\iota_{Full}(O)\\
	U_{2} & = & (\frac{R}{\pi}-1)E[\iota_{Full}(O)\mid O]\\
	U_{3} & = & \frac{R}{\pi}\left(\iota_{Full}(O)-E[\iota_{Full}(O)\mid O]\right)\\
	U_{4} & = & E[\iota_{Full}(O)\mid O]
\end{eqnarray*}

Case 1 - $p$ and $\pi$ are correct
\begin{eqnarray*}
	E\left[U_{1}\right] & = & E[\frac{R}{\pi}\iota_{Full}(O)]\protect\\
	& = & E\left[E[\frac{R}{\pi}\iota_{Full}(O)\mid Y,A,C]\right]\protect\\
	& = & E\left[\frac{\pr(R=1\mid Y,A,C)}{\pr(R=1\mid Y,A,C)}E[\iota_{Full}(O)\mid Y,A,C]\right]\protect\\
	& = & E\left[E[\iota_{Full}(O)\mid O]\right]\protect\\
	& = & 0.
\end{eqnarray*}

Additionally,
\begin{eqnarray*}
	E\left[U_{2}\right] & = & E\left[(\frac{R}{\pi}-1)E[\iota_{Full}(O)\mid O]\right]\\
	& = & E\left[(\frac{R}{\pi}-1)E[\iota_{Full}(O)\mid O]\right]\\
	& = & E[E\left[(\frac{R}{\pi}-1)E[\iota_{Full}(O)\mid O]\mid A,Y,C\right]]\\
	& = & E\left[\left(\frac{\pr(R=1\mid Y,A,C)}{\pr(R=1\mid Y,A,C)}-1\right)E[\iota_{Full}(O)\mid O]\mid A,Y,C]\right]\\
	& = & E\left[0\times E[\iota_{Full}(O)\mid O]\mid A,Y,C]\right]\\
	& = & 0.
\end{eqnarray*}

Case 2 - $m$ and $\pi$ are correct
Follows from Case 1 as $E\left[U_{2}\right]=0$ when $\pi$ is correct
and $E\left[U_{1}\right]=0$ because $E[\iota_{Full}(O)]=0$ when
$m$ is correct.

Case 3 - $p$ and $t$ are correct
\begin{eqnarray*}
	E\left[U_{3}\right] & = & E\left[\frac{R}{\pi}\left(\iota_{Full}(O)-E[\iota_{Full}(O)\mid O]\right)\right]\protect\\
	& = & E\left[E\left[\frac{R}{\pi}\left(\iota_{Full}(O)-E[\iota_{Full}(O)\mid O]\right)\mid O\right]\right]\protect\\
	& = & E\left[\frac{E\left[R=1\mid O\right]}{E\left[\pi\mid O\right]}E\left[\left(\iota_{Full}(O)-E[\iota_{Full}(O)\mid O]\right)\mid O\right]\right]\protect\\
	& = & E\left[\frac{E\left[R=1\vert\mid O\right]}{E\left[\pi\mid O\right]}\times0\right]\protect\\
	& = & 0.
\end{eqnarray*}

Additionally,
\begin{eqnarray*}
	E\left[U_{4}\right] & = & E\left[E[\iota_{Full}(O)\mid O]\right]\\
	& = & E\left[\iota_{Full}(O)\right]\\
	& = & 0.
\end{eqnarray*}

Case 4 - $m$ and $t$ are correct
Follows similarly as Case 3 as $E\left[U_{3}\right]=0$ when $t$
is correct and $E\left[U_{4}\right]=0$ because $E[\iota_{Full}(O)]=0$
when $m$ is correct.

\section*{Example: Problem with the multiply robust method}

The multiply robust method established above requires correct specification
of $p=\pr(A=1\mid L,C)$ or $m\left(Y\mid A,L,C\right)$ and$\pi=\pr(R=1\mid A,Y,C)$
or $t(L\mid A,C,Y)$. The problem with these model specifications
are that they inherently assume we can estimate each model independent
of these others and that they are not related. However, they are closely
related quantities.

For example, suppose $L$ and $A$ were binary. Then
\begin{eqnarray*}
	\logit\left[\pr\left(L=1\mid A,C\right)\right] & = & \phi_{0}^{*}+\phi_{1}^{*}A+\phi_{2}^{*}C
\end{eqnarray*}
and
\begin{eqnarray*}
	\logit\left[\pr\left(A=1\mid L,C\right)\right] & = & \lambda_{0}+\lambda_{1}L+\lambda_{2}C.
\end{eqnarray*}

Under these model specifications, $\phi_{1}^{*}=\lambda_{1}=OR\left(A,L\mid C\right)$.
However we are interested in $\logit\left[\pr\left(L=1\mid A,C,Y\right)\right]=\phi_{0}+\phi_{1}A+\phi_{2}Y+\phi_{3}C$
which will not marginalize over $Y$ to a logistic regression, but
rather a mixture of two logistic regressions for $Y=0$ and $Y=1$.
Therefore we could not specify models for $\logit\left[\pr\left(L=1\mid A,C,Y\right)\right]=\phi_{0}+\phi_{1}A+\phi_{2}Y+\phi_{3}C$
and $\logit\left[\pr\left(A=1\mid L,C\right)\right]=\lambda_{0}+\lambda_{1}L+\lambda_{2}C$
that are compatible with each other. 

It is possible to use the logit link function to model both $t$ and
$p$:{\small{}
	\begin{eqnarray*}
		\logit\left[\pr(L=1\mid A,Y,C)\right] & = & \logit\left[\pr(L=1\mid Y,C)\right]+logOR(L=1,A\mid Y,C)-logE\left[OR(L=1,A\mid Y,C)\mid L=0,Y,C\right]\\
		\logit\left[\pr(A=1\mid L,C)\right] & = & \logit\left[\pr(A=1\mid C)\right]+logOR(A=1,L\mid C)-logE\left[OR(A=1,L\mid C)\mid A=0,C\right].
	\end{eqnarray*}
}{\small \par}

Thus we see that both $t$ and $p$ model the association between
$L$ and $A$, but the former is conditional on $Y$ and $C,$ while
the later is only conditional on $C$. There may not be an intersection
submodel for the particular choice of the nuisance models. In simulation
we can ensure these models are compatible, but in practice we won't
realistically be able to make this assumption. 

Similarly we can relate $m(Y\mid A,C,L)$ and $t(L\mid A,C,Y)$. The
proposed multiply-robust solution assumes we can specify $t(L\mid A,C,Y)$
and $m(Y\mid A,C,L)$ independently of each other. For example suppose
we propose that
\begin{eqnarray*}
	Y\mid A,C,L & \sim & N\left(\nu_{0}+\nu_{1}A+\nu_{2}L^{3}+\nu_{3}C,\sigma_{Y}^{2}\right)\\
	L\mid A,C,Y & \sim & N\left(\phi_{0}+\phi_{1}A+\phi_{2}Y^{2}+\phi_{3}C,\sigma_{L}^{2}\right).
\end{eqnarray*}
Unless $\nu_{2}=0$, these models are not compatible in that there
does not exist a joint distribution for $(L,Y)$ with the given families
as its conditional distributions. Therefore,
for this example, we could never have $t$ and $b$ both be correct
and Case 4 above could never be true.

\section*{Proof of Theorem 1. Detailed Reparametrization of the Likelihood}

We must reparameterize the likelihood because the nuisance parameters
overlap.

\setcounter{theorem}{0}
\begin{lemma}
	\label{lemma1}
	\begin{eqnarray*}
		\frac{f(X_{1}\mid X_{2})}{f(X_{1}=0\mid X_{2})} & = & \int\frac{f(X_{1}\mid X_{2},X_{3})}{f(X_{1}=0\mid X_{2},X_{3})}df(X_{3}\mid X_{2},X_{1}=0).
	\end{eqnarray*}
\end{lemma}

Following \citet{Chen_2007} and \citet{DR_OddsRatio} we define the generalized conditional
odds ratio function of $A$ and $Y$ given $L$ as
\begin{eqnarray*}
	\chi\left(A,Y\mid L\right) & = & \frac{f\left(A\mid Y,L\right)f\left(a_{0}\mid y_{0},L\right)}{f\left(a_{0}\mid Y,L\right)f\left(A\mid y_{0},L\right)}
\end{eqnarray*}
where $\left(a_{0},y_{0}\right)$ is a reference value.

\begin{eqnarray*}
	\frac{f(L\mid Y,A,C)}{f(l_0\mid Y,A,C)} & = & \frac{f(L\mid Y,A,C)}{f(l_0\mid Y,A,C)}\left\{ \frac{f(L\mid A,C)}{f(l_0\mid A,C)}\right\} ^{-1}\frac{f(L\mid A,C)}{f(l_0\mid A,C)}\\
	& = & \frac{f(L\mid Y,A,C)}{f(l_0\mid Y,A,C)}\left\{ \int\frac{f(L\mid y,A,C)}{f(L=0\mid y,A,C)}f(y\mid A,C,l_0)d\mu\left(y\right)\right\} ^{-1}\frac{f(L\mid A,C)}{f(l_0\mid A,C)}\\
	& = & \frac{f(L\mid Y,A,C)}{f(l_0\mid Y,A,C)}\frac{f(l_0\mid y_0,A,C)}{f(L\mid y_0,A,C)}\frac{f(L\mid A,C)}{f(l_0\mid A,C)}\\
	& \times & \left\{ \int\frac{f(L\mid y,A,C)}{f(l_0\mid y,A,C)}\frac{f(l_0\mid y=0,A,C)}{f(L\mid y_0,A,C)}f(y\mid A,C,l_0)d\mu\left(y\right)\right\} ^{-1}\\
	& = & \chi(L,Y\mid A,C)\frac{f(L\mid A,C)}{f(l_0\mid A,C)}\\
	& \times & \left\{ \int\frac{f(L\mid y,A,C)}{f(l_0\mid y,A,C)}\frac{f(l_0\mid y=0,A,C)}{f(L\mid y_0,A,C)}f(y\mid A,C,l_0)d\mu\left(y\right)\right\} ^{-1}\\
	& = & \chi(L,Y\mid A,C)\left\{ \int\chi(L,y\mid A,C)f(y\mid A,C,l_0)d\mu\left(y\right)\right\} ^{-1}\frac{f(L\mid A,C)}{f(l_0\mid A,C)}.
\end{eqnarray*}

Following \citet{Chen_2007} the joint distribution
of $L$ and $Y$ given $A$ and $C$ can be written as
\begin{eqnarray*}
	f(L,Y\mid A,C) & = & \frac{f(L\mid y_0,A,C)\chi(L,Y\mid A,C)f(Y\mid l_0,A,C)}{\int\int\chi\left(l,y\mid A,C\right)f\left(l\mid y_{0},A,C\right)f\left(y\mid l_{0},A,C\right)d\mu\left(l,y\right)}.
\end{eqnarray*}

Then,{\small{}
	\begin{eqnarray*}
		f(L,Y\mid A,C) & = & f(L\mid y_0,A,C)\frac{\chi(L,Y\mid A,C)f(Y\mid l_0,A,C)}{\int\int\chi\left(l,y\mid A,C\right)f\left(l\mid y_{0},A,C\right)f\left(y\mid l_{0},A,C\right)d\mu\left(l,y\right)}\\
		& = & \frac{f(L\mid y_0,A,C)}{f(l_0\mid y_0,A,C)}\frac{\chi(L,Y\mid A,C)f(Y\mid l_0,A,C)}{\left\{ f(l_0\mid y_0,A,C)\right\} ^{-1}\int\int\chi\left(l,y\mid A,C\right)f\left(l\mid y_{0},A,C\right)f\left(y\mid l_{0},A,C\right)d\mu\left(l,y\right)}\\
		& = & \frac{f(L\mid y_0,A,C)}{f(l_0\mid y_0,A,C)}\frac{\chi(L,Y\mid A,C)f(Y\mid l_0,A,C)}{K\left(A,C\right)}\\
		& = & \chi(L,y_0 \mid A,C)\left\{ \int\chi(L,y\mid A,C)f(y\mid A,C,l_0)d\mu\left(y\right)\right\} ^{-1}\frac{f(L\mid A,C)}{f(l_0\mid A,C)}\frac{\chi(L,Y\mid A,C)f(Y\mid l_0,A,C)}{K\left(A,C\right)}\\
		& = & \frac{f\left(L\vert A,C\right)}{f\left(l_0\vert A,C\right)}\left\{ \int\chi(L,y\mid A,C)f(y\mid A,C,l_0)d\mu\left(y\right)\right\} ^{-1}\frac{\chi(L,Y\mid A,C)f(Y\mid l_0,A,C)}{K\left(A,C\right)} \\
		& = & \chi(A,L|C) \frac{f(L |a_0, C)}{f(l_0 | a_0,C)} \left\{ \int\chi(L,y\mid A,C)f(y\mid A,C,l_0)d\mu\left(y\right)\right\} ^{-1}\frac{\chi(L,Y\mid A,C)f(Y\mid l_0,A,C)}{K\left(A,C\right)} \\
		& = &  \frac{\chi(L,Y\mid A,C)f(Y\mid l_0,A,C)}{K\left(A,C\right)}\frac{f(L\mid a_0,C)\chi\left(A,L\mid C\right)}{f(l_0\mid a_0,C)\int\chi(L,y\mid A,C)f(y\mid A,C,l_0)d\mu(y)}
	\end{eqnarray*}
}where $K\left(A,C\right)=\left[f(l_0\mid y_0,A,C)\right]^{-1}\int\int\chi\left(l,y\mid A,C\right)f\left(l\mid y_{0},A,C\right)f\left(y\mid l_{0},A,C\right)d\mu\left(l,y\right)$
and because $\chi(L,y_0 \mid A,C)=1$.

We can see that the joint distribution of $L$ and $Y$ can be expressed
in terms of $f(L\mid a_0,C)$ , $\chi\left(A,L\mid C\right)$, $\chi(L,Y\mid A,C)$,
and $f(Y\mid l_0,A,C)$. As \citet{Chen_2007} shows, these are all variation independent parameters. This allows us to estimate $f(L,Y\mid A,C)$
using maximum likelihood. 

Similarly we can write:
\begin{eqnarray*}
	f(A\mid L,C) & = & \frac{f(A\mid l_0,C)}{f(a_0\mid l_0,C)}\chi(A,L\mid C)\left\{ \int\frac{f(a\mid l_0,C)}{f(a=0\mid l_0,C)}\chi(a,L\mid C)d\mu\left(a\right)\right\} ^{-1}.
\end{eqnarray*}

Thus we see that he propensity score can be expressed in terms of
$\chi\left(A,L\mid C\right)$ and $f(A\mid l_0,C)$ and therefore
both the propensity score and joint distribution of $L$ and $Y$
given $A$ and $C$ require correct specification of $\chi\left(A,L\mid C\right)$.

\section*{Closed Form Estimator}

Recall:
\begin{eqnarray*}
	\iota_{Full}(O) & = & \frac{I(A=0)}{\pr(A=1)}\frac{\pr(A=1\mid L,C)}{\pr(A=0\mid L,C)}(Y-E[Y\mid A=0,L,C])+\frac{I(A=1)}{\pr(A=1)}(E[Y\mid A=0,L,C]-\Psi)
\end{eqnarray*}
where $O=(Y,A,C)$ are the fully observed variables.

Therefore,
\begin{eqnarray*}
	\iota_{Miss}(\Psi) & = & \frac{R}{\pi}\iota_{Full}(\Psi)-(\frac{R}{\pi}-1)E[\iota_{Full}(\Psi)\mid Y,A,C].
\end{eqnarray*}

Thus,
\begin{eqnarray*}
	\iota_{Miss}(\Psi) & = & \frac{R}{\pi}\left\{ \frac{I(A=0)}{\pr(A=1)}\frac{\pr(A=1\mid L,C)}{\pr(A=0\mid L,C)}(Y-E[Y\mid A=0,L,C])+\frac{I(A=1)}{\pr(A=1)}\left(E[Y\mid A=0,L,C]-\Psi\right)\right\} \\
	& - & \left(\frac{R}{\pi}-1\right)\bigg\{ E\bigg[\frac{I(A=0)}{\pr(A=1)}\frac{\pr(A=1\mid L,C)}{\pr(A=0\mid L,C)}(Y-E[Y\mid A=0,L,C])\\
	& + & \frac{I(A=1)}{\pr(A=1)}\left(E[Y\mid A=0,L,C]-\Psi\right)\mid Y,A,C\bigg]\bigg\}\\
	& = & \frac{R}{\pi}\left\{ \frac{I(A=0)}{\pr(A=1)}\frac{\pr(A=1\mid L,C)}{\pr(A=0\mid L,C)}(Y-E[Y\mid A=0,L,C])+\frac{I(A=1)}{\pr(A=1)}E[Y\mid A=0,L,C]\right\} \\
	& - & \frac{R}{\pi}\frac{I(A=1)}{\pr(A=1)}\Psi\\
	& - & \left(\frac{R}{\pi}-1\right)\left\{ \frac{I(A=0)}{\pr(A=1)}YE\left[\frac{\pr(A=1\mid L,C)}{\pr(A=0\mid L,C)}\mid Y,A=0,C\right]\right\} \\
	& + & \left(\frac{R}{\pi}-1\right)\left\{ \frac{I(A=0)}{\pr(A=1)}E\left[\frac{\pr(A=1\mid L,C)}{\pr(A=0\mid L,C)}E[Y\mid A=0,L,C]\mid Y,A=0,C\right]\right\} \\
	& - & \left(\frac{R}{\pi}-1\right)\left\{ \frac{I(A=1)}{\pr(A=1)}E\left[E[Y\mid A=0,L,C]\mid Y,A=1,C]\right]\right\} \\
	& + & \left(\frac{R}{\pi}-1\right)\frac{I(A=1)}{\pr(A=1)}\Psi
\end{eqnarray*}

We set the previous expression equal to zero and solve for $\Psi$. 

Recall that $\frac{R}{\pi}\frac{I(A=1)}{\pr(A=1)}\Psi-(\frac{R}{\pi}-1)\frac{I(A=1)}{\pr(A=1)}\Psi=\Psi\frac{I(A=1)}{\pr(A=1)}$.

Then,
\begin{eqnarray*}
	\Psi\frac{I(A=1)}{\pr(A=1)} & = & \frac{R}{\pi}\left\{ \frac{I(A=0)}{\pr(A=1)}\frac{\pr(A=1\mid L,C)}{\pr(A=0\mid L,C)}(Y-E[Y\mid A=0,L,C])+\frac{I(A=1)}{\pr(A=1)}E[Y\mid A=0,L,C]\right\} \\
	& - & (\frac{R}{\pi}-1)\left\{ \frac{I(A=0)}{\pr(A=1)}YE\left[\frac{\pr(A=1\mid L,C)}{\pr(A=0\mid L,C)}\mid Y,A=0,C\right]\right\} \\
	& + & (\frac{R}{\pi}-1)\left\{ \frac{I(A=0)}{\pr(A=1)}E\left[\frac{\pr(A=1\mid L,C)}{\pr(A=0\mid L,C)}E[Y\mid A=0,L,C]\mid Y,A=0,C\right]\right\} \\
	& - & (\frac{R}{\pi}-1)\left\{ \frac{I(A=1)}{\pr(A=1)}E\left[E[Y\mid A=0,L,C]\mid Y,A=1,C]\right]\right\} .
\end{eqnarray*}

We can then look at each term on the right hand side of the equation
separately and consider the models proposed in the main body of the
paper.

Let:
\begin{eqnarray*}
	V_{1} & = & \frac{R}{\pi}\left\{ \frac{I(A=0)}{\pr(A=1)}\frac{p}{1-p}(Y-m\left(0,L,C\right))+\frac{I(A=1)}{\pr(A=1)}m\left(0,L,C\right)\right\} \\
	V_{2} & = & (\frac{R}{\pi}-1)\left\{ \frac{I(A=0)}{\pr(A=1)}YE\left[\frac{p}{1-p}\mid Y,A=0,C\right]\right\} \\
	V_{3} & = & (\frac{R}{\pi}-1)\left\{ \frac{I(A=0)}{\pr(A=1)}E\left[\frac{p}{1-p}E[Y\mid A=0,L,C]\mid Y,A=0,C\right]\right\} \\
	V_{4} & = & (\frac{R}{\pi}-1)\left\{ \frac{I(A=1)}{\pr(A=1)}E\left[m\left(0,L,C\right)\mid Y,A=1,C]\right]\right\} .
\end{eqnarray*}

$V_{1}$ requires models for $\pi$, $p$, and for $m\left(0,L,C\right)=\mu_{Y}^{0}$,
which are easily estimated as described in the main body of the paper.

$V_{2}$ also requires models for $\pi$ and $p$ in addition to $t\left(A,Y,C\right)$.
Using that fact that, for $X\sim N(\mu,\sigma^{2})$, the moment generating
function is $E[e^{tX}]=e^{\mu t+\frac{1}{2}\sigma^{2}t^{2}}$, we
have that:
\begin{eqnarray*}
	E\left[\frac{p}{1-p}\mid Y,A=0,C\right] & = & E\left[e^{\lambda_{0}+\lambda_{1}L+\lambda_{2}C}\mid Y,A=0,C\right]\\
	& = & e^{\lambda_{0}+\lambda_{2}C}E\left[e^{\lambda_{1}L}\mid Y,A=0,C\right]\\
	& = & e^{\lambda_{0}+\lambda_{2}C}e^{\lambda_{1}\mu_{L}^{0}+\frac{1}{2}\sigma_{L}^{2}\lambda_{1}^{2}}.
\end{eqnarray*}

Thus $V_{2}$ can be expressed as $(\frac{R}{\pi}-1)\left\{ \frac{I(A=0)}{\pr(A=1)}Ye^{\lambda_{0}+\lambda_{2}C+\lambda_{1}\mu_{L}^{0}+\frac{1}{2}\sigma_{L}^{2}\lambda_{1}^{2}}\right\} $.

$V_{3}$ requires models for $\pi$, $p$, $t\left(0,Y,C\right)$
and $m\left(0,L,C\right)$. Let $E\left[\frac{p}{1-p}m\left(0,L,C\right)\mid Y,A=0,C\right]=\zeta$.
Then,
\begin{eqnarray*}
	\zeta & = & E\left[e^{\lambda_{0}+\lambda_{1}L+\lambda_{2}C}(\nu_{0}+\nu_{2}L+\nu_{3}C)\mid Y,A=0,C\right]\\
	& = & e^{\lambda_{0}+\lambda_{2}C}\left(\nu_{0}+\nu_{3}C\right)e^{\lambda_{1}\mu_{L}^{0}+\frac{1}{2}\sigma_{L}^{2}\lambda_{1}^{2}}\\
	& + & \nu_{2}e^{\lambda_{0}+\lambda_{2}C}\left(\mu_{L}^{0}+\sigma_{L}^{2}\lambda_{1}\right)e^{\lambda_{1}\mu_{L}^{0}+\frac{1}{2}\sigma_{L}^{2}\lambda_{1}^{2}}.
\end{eqnarray*}

Therefore we can see that $V_{3}$ can be expressed as

{\footnotesize{}$(\frac{R}{\pi}-1)\left\{ \frac{I(A=0)}{\pr A=1)}\left[\left(\nu_{0}+\nu_{3}C\right)e^{\lambda_{0}+\lambda_{2}C+\lambda_{1}\mu_{L}^{0}+\frac{1}{2}\sigma_{L}^{2}\lambda_{1}^{2}}+\nu_{2}\left(\mu_{L}^{0}+\sigma_{L}^{2}\lambda_{1}\right)e^{\lambda_{0}+\lambda_{2}C+\lambda_{1}\mu_{L}^{0}+\frac{1}{2}\sigma_{L}^{2}\lambda_{1}^{2}}\right]\right\} $}{\footnotesize \par}

$V_{4}$ requires models for $\pi$, $t\left(A,Y,C\right)$ and $m\left(0,L,C\right)$.
\begin{eqnarray*}
	E\left[E[Y\mid A=0,L,C]\mid Y,A=1,C]\right] & = & E\left[\nu_{0}+\nu_{2}L+\nu_{3}C\mid Y,A=1,C]\right]\\
	& = & \nu_{0}+\nu_{3}C+\nu_{2}E\left[L\mid Y,A=1,C]\right]\\
	& = & \nu_{0}+\nu_{3}C+\nu_{2}\left(\phi_{0}+\phi_{1}+\phi_{2}Y+\phi_{3}C\right)
\end{eqnarray*}

Thus, in our example, $V_{4}$ can be expressed as $(\frac{R}{\pi}-1)\left\{ \frac{I(A=1)}{\pr(A=1)}\nu_{0}+\nu_{3}C+\nu_{2}\left(\phi_{0}+\phi_{1}+\phi_{2}Y+\phi_{3}C\right)\right\} $

To estimate $\Psi$, we calculate the sum of the four terms for each
subject, then take the sample mean across all subjects. 

\end{document}